\documentclass[journal]{IEEEtran}
\usepackage{amsmath}
\usepackage{amssymb}
\usepackage{amsthm}
\usepackage{array}
\usepackage{booktabs}
\usepackage{cite}
\usepackage{color}
\usepackage{xcolor}
\usepackage{caption}
\usepackage{epsfig,latexsym}
\usepackage{float}
\usepackage{fancyhdr}
\usepackage{graphicx}
\usepackage{indentfirst}
\usepackage{mdwmath}
\usepackage{mdwtab}
\usepackage{subfigure}
\usepackage{setspace}
\usepackage{times}
\usepackage{url}
\usepackage{multirow}
\usepackage{algorithm}
\usepackage{algorithmic}
\usepackage{verbatim}
\usepackage{algorithm,algorithmic}
\usepackage{mathrsfs}

\newtheorem{remark}{Remark}
\newtheorem{theorem}{Theorem}

\newtheorem{lemma}{Lemma}

\newtheorem{corollary}{Corollary}

\begin{document}

\title{\huge{Uplink Positioning for PASS in Multipath Environments}}

\author{Yaoyu Zhang, Xin Sun, Tianwei Hou,~\IEEEmembership{Member,~IEEE}, Anna Li,~\IEEEmembership{Member,~IEEE}, \\and Yuanwei Liu,~\IEEEmembership{Fellow,~IEEE}

\thanks{This work was supported in part by the Beijing Natural Science Foundation L232041, and in part by EPSRC grant numbers to acknowledge are EP/W004100/1, EP/W034786/1 and EP/Y037243/1.}       
\thanks{Yaoyu Zhang and Xin Sun are with the School of Electronic and Information Engineering, Beijing Jiaotong University, Beijing 100044, China (e-mail: yaoyu.zhang@bjtu.edu.cn; xsun@bjtu.edu.cn).}
\thanks{Tianwei Hou is with the Beijing Key Laboratory of Convergent Communications and Networking Technologies in LEO Satellite Systems, and also with the State Key Laboratory of Networking and Switching Technology, Beijing University of Posts and Telecommunications, Beijing 100876, China (email: htw@bupt.edu.cn). (Corresponding author)}  
\thanks{Anna Li is with the School of Computing and Communications, Lancaster University, Lancaster LA1 4WA, U.K. (e-mail: a.li16@lancaster.ac.uk). }
\thanks{Yuanwei Liu is with the Department of Electrical and Electronic Engineering, The University of Hong Kong, Hong Kong (e-mail: yuanwei@hku.hk). }}

\maketitle

\begin{abstract}
Pinching-antenna systems (PASS) enhance wireless propagation by activating or placing pinching antennas (PAs) near users. Therefore, accurate uplink positioning is essential for efficient communication. In this paper, an uplink multi-carrier positioning framework is established for PASS in multipath environments. Matrix pencil (MP)-based and low-complexity Rank-1 ranging algorithms are proposed to estimate the distances between the PAs and the user. For the MP-based ranging algorithm, the line-of-sight (LoS) component is separated from non-line-of-sight components by exploiting the shift-invariance property of the Hankel matrix, thereby enabling accurate distance estimation. For the Rank-1 ranging algorithm, the dominant LoS delay is directly isolated through truncated singular value decomposition, thereby avoiding matrix inversions. Subsequently, a two-stage weighted nonlinear least-squares (WNLS) positioning algorithm is designed to estimate the three-dimensional user position. To gain further insights, a comprehensive theoretical performance analysis of the proposed ranging and positioning algorithms is conducted. The closed-form ranging variances and position error bound (PEB) are derived to reveal the error propagation mechanism. Numerical results demonstrate that: i) The MP-based algorithm achieves higher accuracy and robustness than the Rank-1-based algorithm, while the Rank-1-based algorithm has lower computational complexity. ii) The positioning error of the MP-based algorithm follows the same trend as the derived PEB, whereas the Rank-1 algorithm exhibits an error floor due to multipath bias. iii) The positioning accuracy of the MP algorithm improves as the number of subcarriers increases.
\end{abstract}


\section{Introduction}

Multiple-input multiple-output has become a key technology for future wireless networks, since large antenna arrays can provide spatial multiplexing, beamforming, and diversity gains for both communication and localization applications~\cite{MIMO}. However, conventional fixed-position antenna arrays usually require dense antenna deployment and dedicated radio-frequency chains, resulting in high hardware cost and limited deployment flexibility. To overcome these limitations, reconfigurable antenna technologies have recently attracted increasing attention. Reconfigurable intelligent surfaces (RIS) can reshape the wireless propagation environment and establish virtual line-of-sight (LoS) links by controlling passive reflecting elements~\cite{RIS_1},~\cite{RIS_2}. Nevertheless, the cascaded transmitter-RIS-user link suffers from severe double-fading path loss. More recently, fluid antenna systems (FAS) and movable antennas (MA) have further enhanced channel reconfigurability by changing antenna positions~\cite{FAS_1, FAS_2, MA, MA2}. However, the movement regions of FAS and MA are usually limited to the wavelength scale, which restricts their ability to provide large-scale channel reconfiguration, especially in millimeter-wave bands. In this context, pinching-antenna systems (PASS) have emerged as a promising alternative, offering meter-scale antenna reconfigurability through low-cost dielectric waveguides and pinching antennas (PAs)~\cite{PASS_1}.

A PASS typically consists of a dielectric waveguide connected to an access point (AP) and multiple PAs deployed along the waveguide. By attaching a separate dielectric to a selected position on the waveguide, a PA is formed, which perturbs the local guided mode and enables coupling between the guided wave and the free-space wave. As a result, each activated PA functions as a reconfigurable radiation/reception point, enabling flexible antenna-position adjustment according to the service demand~\cite{PASS_2, PASS_3, PASS_4}. Owing to this unique architecture, PASS has attracted increasing attention in wireless communications. For instance, Tegos et al. maximized the minimum data rate of uplink PASS by jointly optimizing PA positions and resource allocation, showing that flexible PA deployment can improve both throughput and fairness~\cite{PASS_com1}. Bereyhi et al. extended PASS to downlink multiuser MIMO systems and jointly designed the digital precoder and activated PA locations, demonstrating significant weighted sum-rate gains over fixed-position antenna systems~\cite{PASS_com2}. Lv et al. developed beam training schemes, which reduce the training overhead while maintaining beam alignment accuracy~\cite{PASS_com3}. To provide theoretical insights, Tyrovolas et al. developed a comprehensive analytical framework for PASS, deriving closed-form expressions for the outage probability and average rate~\cite{PASS_the}. The analysis further characterized the optimal PA placement under lossy waveguides, showing that PASS outperforms conventional fixed-antenna systems in reliability and data rate. Hou et al. analyzed the uplink performance for PASS, deriving closed-form and asymptotic ergodic-rate expressions~\cite{PASS_com4}. In another work, Hou et al. further investigated the antenna gain of PA arrays, offering a more practical array-level implementation of PASS~\cite{PASS_com5}. To support PA deployment and beamforming in practice, channel estimation methods for PASS were proposed to acquire channel state information over candidate PA positions~\cite{PASS_channel},~\cite{PASS_channel2}.

Despite the demonstrated potential of PASS in enhancing communication performance, its positioning capabilities remain largely unexplored. Most existing works focus exclusively on communication-oriented designs, which typically rely on the assumption of known user positions to optimize PA deployment. Therefore, accurate uplink positioning is a fundamental prerequisite for fully exploiting the communication gains of PASS. Moreover, considering its geometrically deterministic channel model and meter-scale reconfigurability, PASS is particularly promising for indoor positioning. 

Several recent studies have investigated PASS-based positioning. Zhang et al. first proposed a PASS-based indoor positioning framework under a single-waveguide and LoS setting, where received signal strength (RSS)-based ranging and WLS algorithm were adopted to estimate the user position~\cite{PASS_poi_1}. This work provides an initial benchmark for PASS localization, but it relies only on RSS. To exploit richer signal information, Feng et al. developed a phase-aware localization framework by jointly using the amplitude and phase of the received complex signal~\cite{PASS_poi_2}. From a network-level perspective, He et al. investigated RSS-based PA-assisted localization using stochastic geometry~\cite{PASS_poi_3}. They derived the target localizability and Cramer-Rao lower bound distribution, revealing the impact of random waveguide deployment. To improve geometric diversity, Zhang et al. further extended PASS positioning to multi-waveguide scenarios, including multi-waveguide single-PA and multi-waveguide multi-PA configurations, together with an analysis of the corresponding integrated positioning and communication performance~\cite{PASS_poi_41,PASS_poi_4}. Beyond pure LoS positioning, Xu et al. proposed a joint user localization and channel estimation framework based on multiple PA subarrays, where an orthogonal matching pursuit-based geometry-consistent localization algorithm was developed to estimate user/scatterer locations~\cite{PASS_poi_5}. However, this method relies on dictionary-based sparse recovery, which may suffer from grid mismatch. More recently, Liu et al. investigated channel estimation and localization in uplink PASS-OFDM systems, showing that multipath-induced delay dispersion can be exploited for user localization~\cite{PASS_poi_6}. Nevertheless, this framework is based on iterative Bayesian inference, leading to relatively high algorithmic complexity. The low-complexity positioning algorithms and the corresponding error mechanism analysis remain insufficiently investigated.

\subsection{Motivation and contribution}

Although PASS have been extensively studied for communication performance analysis and optimization, its positioning capability remains insufficiently explored. Moreover, most communication-oriented PASS studies assume that the user position is available when optimizing PA deployment, beamforming, or resource allocation, overlooking the critical prerequisite of obtaining precise user locations in practice. Furthermore, while a few recent studies have explored PASS-based positioning, they are primarily restricted to pure LoS scenarios. In practical indoor environments where multipath propagation is inevitable, existing methods addressing non-line-of-sight (NLoS) components rely on dictionary-based sparse recovery or iterative Bayesian inference. These approaches may suffer from grid mismatch and high computational complexity. Consequently, the development of low-complexity positioning algorithms for PASS and the rigorous analysis of error mechanisms in multipath environments remain insufficiently investigated.

To overcome these limitations, this paper investigates uplink positioning techniques for PASS in practical multipath environments. The main theoretical and algorithmic contributions of this paper are outlined below:

\begin{itemize}
  \item We propose a comprehensive uplink multi-carrier positioning framework for PASS in practical multipath environments. Based on the structure of the received signals, we propose a matrix pencil (MP)-based ranging algorithm to extract the multipath delays from the channel observations. By exploiting the shift-invariance property of the Hankel matrix, the proposed MP-based ranging algorithm separates the LoS component from NLoS components, estimating the LoS distance for subsequent positioning.        
  \item To further reduce the computational complexity, we develop a low-complexity Rank-1 ranging algorithm. Instead of performing full subspace decomposition to resolve all multipath components, this algorithm employs a truncated singular value decomposition (SVD) to obtain the optimal Rank-1 approximation of the Hankel matrix. By exploiting the shift-invariance property, the proposed algorithm directly extracts the dominant LoS delay via a least squares (LS) criterion. Therefore, the Rank-1-based algorithm avoids pseudo-inverse computation and eigenvalue decomposition, providing a lightweight alternative for practical PASS positioning. Subsequently, based on the extracted LoS distances, a two-stage weighted nonlinear least-squares (WNLS) is utilized to robustly estimate the three-dimensional (3D) user position for practical PASS deployments.
  \item To gain further insights, we provide a theoretical performance analysis for the proposed ranging and positioning algorithms. Specifically, the ranging error variances of the MP-based and Rank-1-based algorithms are derived under high-SNR approximations. Based on these ranging variances, the Fisher information matrix (FIM) of the positioning model is established, and the corresponding positioning error bound (PEB) is obtained. Our analysis reveals the propagation of ranging errors into the final 3D positioning errors, highlighting the impacts of number of subcarriers, bandwidth and geometric dilution of precision (GDoP).  
  \item The simulation results confirm our analysis, and the results also demonstrate that: i) The MP-based algorithm achieves higher ranging and positioning accuracy in multipath environments, while the Rank-1-based algorithm exhibits an error floor due to residual multipath bias. ii) Increasing the number of subcarriers improves the MP-based positioning accuracy, whereas the Rank-1-based algorithm is mainly limited by deterministic NLoS-induced bias. iii) The MP-based algorithm achieves lower positioning errors over the room, and shows stronger robustness to the variations of user height. iv) The simulated positioning error of the MP-based algorithm follows the same trend as the derived PEB, while the Rank-1-based algorithm trades positioning accuracy for lower computational complexity.
\end{itemize}

\subsection{Organization and Notations}

The remainder of this paper is organized as follows. Section~\uppercase\expandafter{\romannumeral2} presents the system model for uplink PASS positioning in multipath environments. Section~\uppercase\expandafter{\romannumeral3} develops the proposed ranging and positioning algorithms, including the MP-based ranging algorithm, the low-complexity Rank-1 Hankel approximation ranging algorithm and the two-stage WNLS positioning algorithm. Section~\uppercase\expandafter{\romannumeral4} provides the theoretical performance analysis, where the ranging error variances, the FIM, and the PEB are derived. In Section~\uppercase\expandafter{\romannumeral5}, we discuss the numerical results of the proposed algorithms. Finally, Section~\uppercase\expandafter{\romannumeral6} concludes this paper. 

Throughout this paper, lowercase italic letters are used for scalars, lowercase bold upright letters for vectors, and uppercase bold upright letters for matrices. The notation $\left| {\cdot} \right|$, $\left\| \cdot \right\|_2$ and $\left\| \cdot \right\|_F$ denote the absolute value or modulus, Euclidean norm, and Frobenius norm, respectively. ${\rm{Tr}}\left( \cdot \right)$ denotes the trace of the matrix. ${\left( {\cdot} \right)^t}$ is the $t$-th iteration of the algorithm. The superscript ${\left( {\cdot} \right)^ * }$, ${\left( {\cdot} \right)^ T }$ and ${\left( {\cdot} \right)^ H }$ denote the conjugate, transpose, and conjugate transpose, respectively. ${\left( {\cdot} \right)^ {-1} }$ and ${\left( {\cdot} \right)^ {\dagger} }$ represent the matrix inverse and pseudo-inverse, respectively.

\section{System Model}

\begin{figure}[t!]
\centering
\includegraphics[width =3.5in]{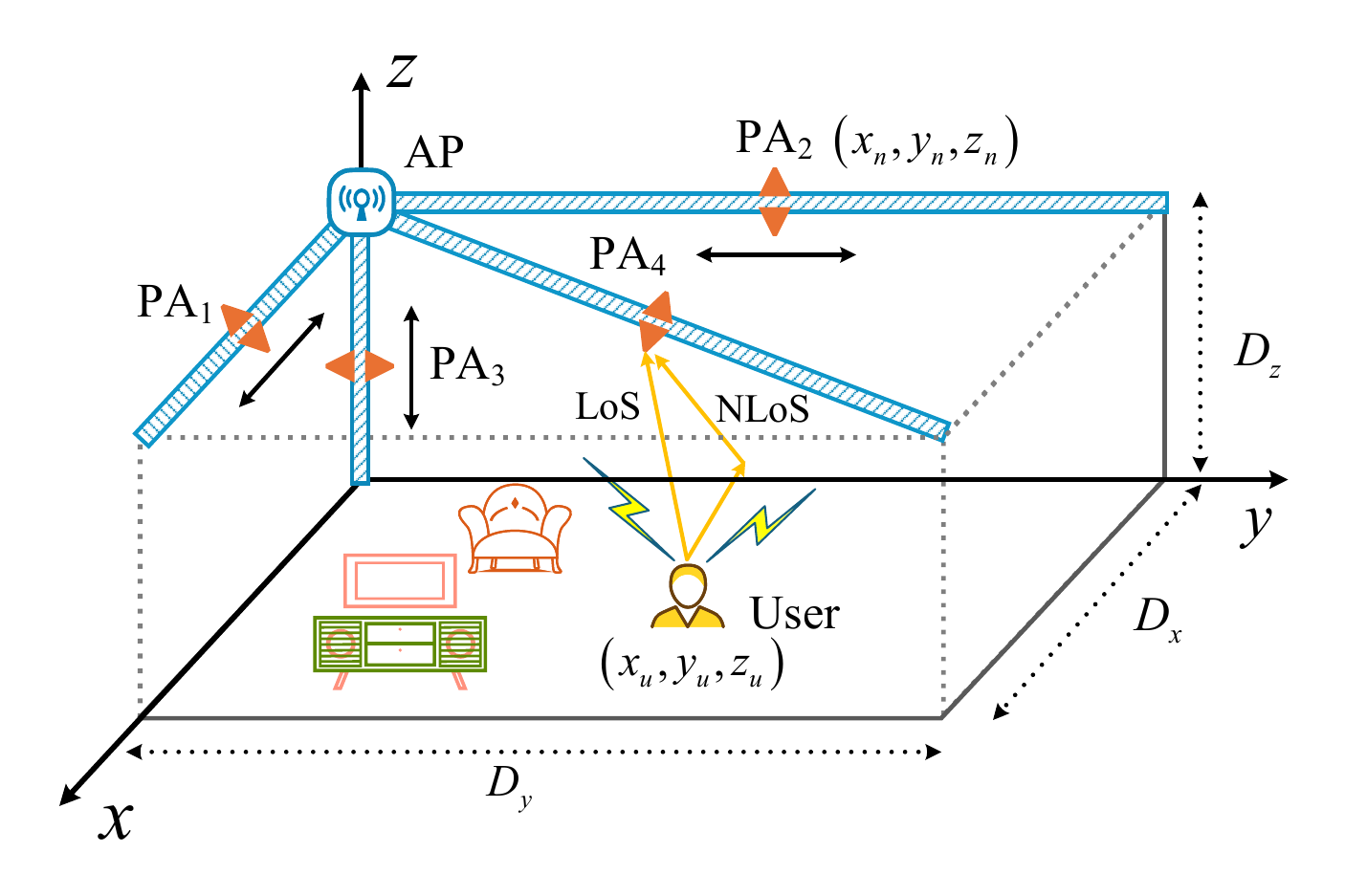}
\caption{PASS-based uplink positioning system model.}
\label{system model}
\end{figure}

\subsection{Antenna Model}

As shown in Fig.~\ref{system model}, we consider a 3D indoor positioning framework for PASS in multipath environments. The room dimensions are bounded by ${D_x}$, ${D_y}$ and ${D_z}$ along the $x$-axis, $y$-axis and $z$-axis, respectively. An access point (AP) is deployed at the corner of the ceiling, whose position is defined as $\left( {0,0,{D_z}} \right)$, connecting to four waveguides that extend along the $x$-axis, $y$-axis, $z$-axis and the diagonal of the ceiling. The positions of four waveguides are defined as ${W_1} = \left\{ {\left( {x,0,{D_z}} \right)\mid 0 \le x \le {D_x}} \right\}$, ${W_2} = \left\{ {\left( {0,y,{D_z}} \right)\mid 0 \le y \le {D_y}} \right\}$, ${W_3} = \left\{ {\left( {0,0,z} \right)\mid 0 \le z \le {D_z}} \right\}$ and ${W_4} = \left\{ {\left( {x,\frac{{{D_y}}}{{{D_x}}}x,{D_z}} \right)\mid 0 \le x \le {D_x}} \right\}$, respectively. Each waveguide is equipped with a single PA, which can move along the waveguide. Let $\left( {{x_n},{y_n},{z_n}} \right)$ and $\left( {{x_u},{y_u},{z_u}} \right)$ be the coordinates of the PA and the single-antenna user, respectively, where $n = 1,2,3,4$. The positions of four PAs are considered to be known.

\subsection{Channel Model}

In this paper, we consider a sparse millimeter-wave (mmWave) channel, where the scattering is expected to be very limited. Measurement campaigns conducted in NLoS environments have demonstrated that mmWave channels typically exhibit only 3 to 4 scattering clusters~\cite{scatter}. The mmWave channel from the user to the $n$-th PA can be modelled as:
\begin{equation}\label{channel model}
\begin{aligned}
h_{nk} = \sum\limits_{l = 1}^L {{\alpha _{nl}}{e^{ - j2\pi \left( {{f_0} + k\Delta f} \right){\tau _{nl}}}}},
\end{aligned}
\end{equation}
where $k \in \left\{ {0,1, \ldots ,K - 1} \right\}$ denotes the index of $K$ subcarriers, and $l$ represents the path index among $L$ multipath components. $f_0$ and $\Delta f$ represent the carrier frequency and subcarrier spacing, respectively. ${\tau _{nl}}$ is the path delay, which can be written as:
\begin{equation}\label{path delay}
\begin{aligned}
{\tau _{nl}} = \frac{{{d_{nl}}}}{c},
\end{aligned}
\end{equation}
where $c$ represents the speed of light, and $d_{nl}$ denotes the total distance from the user to the $n$-th PA on the $l$-th path. The distance of the LoS path can be expressed as:
\begin{equation}\label{distance of the LoS path}
\begin{aligned}
{d_{n1}} = \sqrt {{{\left( {{x_u} - {x_n}} \right)}^2} + {{\left( {{y_u} - {y_n}} \right)}^2} + {{\left( {{z_u} - {z_n}} \right)}^2}}.
\end{aligned}
\end{equation}

We define ${\alpha _{nl}}$ as the path gain coefficient. The path gain coefficient of LoS and NLoS components can be respectively written as: 
\begin{equation}\label{LoS component}
\begin{aligned}
{\alpha _{n1}} = {\frac{\lambda }{{4\pi {d_{n1}}}}},
\end{aligned}
\end{equation}
and
\begin{equation}\label{NLoS component}
\begin{aligned}
{\alpha _{nl}} = {\frac{\lambda }{{{{\left( {4\pi } \right)}^{\frac{3}{2}}}{d_{us}}{d_{sn}}}}},
\end{aligned}
\end{equation}
where $d_{us}$ and $d_{sn}$ represent the distance from the user to scatterer and the scatterer to the $n$-th PA, respectively, which satisfy the relationship ${d_{us}} + {d_{sn}} = {d_{nl}}$. $\lambda$ is the wavelength of the signal.

The signals are radiated into the waveguides through PAs. As the signals propagate along the waveguides, the amplitude attenuation and phase shift per-meter can be characterized by the complex propagation constant as follows~\cite{Microwave_engineering}:
\begin{equation}\label{propagation constant}
\gamma  = \frac{{2\pi \sqrt {{\varepsilon _r}} }}{\lambda }\left( {\frac{{\tan \delta }}{2} + j} \right),
\end{equation}
where ${{\varepsilon _r}}$ denotes relative permittivity of the waveguide, and ${\tan \delta }$ represents the loss angle tangent. The waveguide channel can be given by:
\begin{equation}\label{waveguide channel}
\begin{aligned}
h_{w,n} = \exp \left( { - \frac{{2\pi \sqrt {{\varepsilon _r}} }}{\lambda }\left( {\frac{{\tan \delta }}{2} + j} \right){d_{An}}} \right),
\end{aligned}
\end{equation}
where $d_{An}$ represents the distance between the $n$-th PA and AP, which can be written as:
\begin{equation}\label{PA to AP}
{d_{An}} = \sqrt {x_n^2 + y_n^2 + {{\left( {{z_n} - {D_z}} \right)}^2}}.
\end{equation}

\subsection{Signal Model}

The user transmits $I$ pilot symbols, each symbol modulated by $K$ subcarriers. The $i$-th pilot signal received by AP from the $n$-th waveguide can be expressed as:
\begin{equation}\label{signal}
\begin{aligned}
{y_{nk,i}} = \sqrt{P}h_{w,n}{h_{nk}}{x_i} + {n_{ik}},
\end{aligned}
\end{equation}
where $x_i$ denotes the $i$-th pilot symbol, and ${n_{ik}}$ is the zero-mean additive white Gaussian noise (AWGN) with variance $\sigma^2$. $P$ denotes the signal transmit power.

\section{Ranging and positioning Algorithm}

\subsection{Matrix Pencil-based Ranging Algorithm}

As shown in \eqref{channel model}, the channel from the user to PA implicitly contains the delay information, which can be exploited for ranging. Since the coordinates of PAs and the pilot symbols are considered to be known, the channel from the user to PA can be estimated from the received signals, which can be written as:
\begin{equation}\label{channel}
\begin{aligned}
{\hat h_{nk,i}} = \frac{{{y_{nk,i}}\exp \left( {j\frac{{2\pi \sqrt {{\varepsilon _r}} {d_{An}}}}{\lambda }} \right)}}{{\sqrt P {x_i}\exp \left( { - \frac{{\pi \sqrt {{\varepsilon _r}} \tan \delta }{d_{An}}}{\lambda }}  \right)}} = {h_{nk}} + n_{ik}^{'},
\end{aligned}
\end{equation}
where $n_{ik}^{'} = \frac{{{n_{ik}}\exp \left( {j\frac{{2\pi \sqrt {{\varepsilon _r}} {d_{An}}}}{\lambda }} \right)}}{{\sqrt P {x_i}\exp \left( { - \frac{{\pi \sqrt {{\varepsilon _r}} \tan \delta }{d_{An}}}{\lambda }} \right)}}$.

Furthermore, the channel ${\hat h_{nk,i}}$ can be written as a sum of exponents:
\begin{equation}\label{rewritten channel}
\begin{aligned}
{{\hat h}_{nk,i}} &= \sum\limits_{l = 1}^L \underbrace{{\alpha _{nl}}{e^{ - j2\pi {f_0}{\tau _{nl}}}}}_{\beta _{nl}} {\left( \underbrace{{{e^{ - j2\pi \Delta f{\tau _{nl}}}}}}_{z_{nl}} \right)^k} + n_{ik}^{'}\\
&= \sum\limits_{l = 1}^L {{\beta _{nl}}{{\left( {{z_{nl}}} \right)}^k}} + n_{ik}^{'}.
\end{aligned}
\end{equation}

\begin{algorithm}[!ht]
    \renewcommand{\algorithmicrequire}{\textbf{Input:}}
	\renewcommand{\algorithmicensure}{\textbf{Output:}}
	\caption{MP-Based Ranging Algorithm}
    \label{alg:MP_ranging}
    \begin{algorithmic}[1] 
        \REQUIRE  Averaged channel observation $\bar{h}_{nk}$, waveguide parameters, subcarrier spacing $\Delta f$, and unit circle tolerance $\epsilon$. 
	    \ENSURE The estimated LoS distance $\hat {d}_{n1}$. 
        \STATE \textbf{Matrix construction:} Select the pencil parameter $M$ and construct the Hankel matrices $\bar{\mathbf{H}}_0$ and $\bar{\mathbf{H}}_1$ based on $\bar{h}_{nk}$.
        \STATE \textbf{SVD:} Perform SVD on $\bar{\mathbf{H}}_0$ to determine the effective number of multipath components $\hat{L}$. 
        \STATE \textbf{Subspace projection:} Extract the signal subspace $\mathbf{U}_s = \mathbf{U}(:, 1:\hat{L})$, and then project the Hankel matrices onto ${\bf U}_s$ to obtain the robust matrices ${\mathbf{H}}_{0s}$ and ${\mathbf{H}}_{1s}$.
        \STATE \textbf{Pole extraction:} Compute the signal poles $\hat{z}_{nl}$ by calculating the eigenvalues of $ \mathbf{H}_{1s} \mathbf{H}_{0s}^\dagger$.
        \STATE \textbf{Pole filtering:} Filter out the spurious poles caused by noise by using the unit circle criterion $\big| |\hat{z}_{nl}| - 1 \big| \le \epsilon$.
        \STATE \textbf{Delay estimation:} Estimate the multipath delays $\hat{\tau}_{nl}$ by extracting the phase angles.
        \STATE \textbf{LoS distance calculation:} Identify the shortest estimated delay $\hat{\tau}_{n1}$ from the valid paths and calculate the LoS distance $\hat{d}_{n1} = c\hat{\tau}_{n1}$.
        \RETURN The estimated LoS distance $\hat{d}_{n1}$
    \end{algorithmic}
\end{algorithm}

Thus, we can employ MP method to estimate the LoS delays in the channel observation, which can be subsequently utilized to determine the user's position~\cite{matrix_pencil}. Algorithm~\ref{alg:MP_ranging} outlines the procedure of MP-based ranging. First, to improve the effective signal-to-noise ratio (SNR), we perform coherent averaging to the channel observation $\hat{h}_{nk,i}$. By taking the first pilot as the reference, the relative phase offset of the $i$-th pilot is estimated via cross-correlation, which can be written as:
\begin{equation}\label{eq:phase_offset}
    \hat{\theta}_i = \angle \left( \sum_{k=0}^{K-1} \hat{h}_{nk,i} \hat{h}_{nk,1}^* \right).
\end{equation}
By compensating for this phase drift, the coherent averaged channel observation $\bar{h}_{nk}$ is given by:
\begin{equation}\label{eq:averaged_channel}
    \bar{h}_{nk} = \frac{1}{I} \sum_{i=1}^{I} \hat{h}_{nk,i} e^{-j\hat{\theta}_i},
\end{equation}
where $I$ represents the number of pilot symbols.

For a fixed PA, we select the pencil parameters $M$ and $G$ to construct the Hankel matrices, with $G = K - M$ and $L \ll \min \left( {M,G} \right)$. According to~\cite{matrix_pencil}, the Hankel matrices can be written as:
\begin{equation}\label{Hankel matrices H0}
{{\bf{\bar H}}_0} = \left[ {\begin{array}{*{20}{c}}
{{{\bar h}_{n0}}}&{{{\bar h}_{n1}}}& \cdots &{{{\bar h}_{n\left( {G - 1} \right)}}}\\
{{{\bar h}_{n1}}}&{{{\bar h}_{n2}}}& \cdots &{{{\bar h}_{nG}}}\\
 \vdots & \vdots & \ddots & \vdots \\
{{{\bar h}_{n\left( {M - 1} \right)}}}&{{{\bar h}_{nM}}}& \cdots &{{{\bar h}_{n\left( {K - 2} \right)}}}
\end{array}} \right],
\end{equation}

\begin{equation}\label{Hankel matrices H1}
{{\bf{\bar H}}_1} = \left[ {\begin{array}{*{20}{c}}
{{{\bar h}_{n1}}}&{{{\bar h}_{n2}}}& \cdots &{{{\bar h}_{nG}}}\\
{{{\bar h}_{n2}}}&{{{\bar h}_{n3}}}& \cdots &{{{\bar h}_{n\left( {G + 1} \right)}}}\\
 \vdots & \vdots & \ddots & \vdots \\
{{{\bar h}_{nM}}}&{{{\bar h}_{n\left( {M + 1} \right)}}}& \cdots &{{{\bar h}_{n\left( {K - 1} \right)}}}
\end{array}} \right].
\end{equation}

\subsubsection{Ideal Noise-Free Case Analysis} 
To investigate the fundamental mechanism of the MP method, we first consider the ideal noise-free scenario. The Hankel matrices in noise-free scenario can be respectively factorized as:
\begin{equation}\label{H0}
{{\bf{H}}_0} = {{\bf{Z}}_M}{\bf{BZ}}_G^T,
\end{equation}
and
\begin{equation}\label{H1}
{{\bf{H}}_1} = {{\bf{Z}}_M}{\bf{BZZ}}_G^T,
\end{equation}
where
\begin{equation}\label{ZM}
{{\bf{Z}}_M} = \left[ {\begin{array}{*{20}{c}}
1&1& \cdots &1\\
{{z_{n1}}}&{{z_{n2}}}& \cdots &{{z_{nL}}}\\
 \vdots & \vdots & \ddots & \vdots \\
{z_{n1}^{M - 1}}&{z_{n2}^{M - 1}}& \cdots &{z_{nL}^{M - 1}}
\end{array}} \right],
\end{equation}

\begin{equation}\label{B}
{\bf{B}} = {\rm{diag}}\left( {{\beta _{n1}},{\beta _{n2}}, \ldots ,{\beta _{nL}}} \right),
\end{equation}

\begin{equation}\label{ZN}
{{\bf{Z}}_G} = \left[ {\begin{array}{*{20}{c}}
1&1& \cdots &1\\
{{z_{n1}}}&{{z_{n2}}}& \cdots &{{z_{nL}}}\\
 \vdots & \vdots & \ddots & \vdots \\
{z_{n1}^{G - 1}}&{z_{n2}^{G - 1}}& \cdots &{z_{nL}^{G - 1}}
\end{array}} \right],
\end{equation}

\begin{equation}\label{Z}
{\bf{Z}} = {\rm{diag}}\left( {{z_{n1}},{z_{n2}}, \ldots ,{z_{nL}}} \right).
\end{equation}

Based on the factorizations in \eqref{H0} and \eqref{H1}, we can construct a matrix pencil ${{\bf{H}}_1} - \mu {{\bf{H}}_0}$, which can be expressed as:
\begin{equation}\label{matrix pencil}
{{\bf{H}}_1} - \mu {{\bf{H}}_0} = {{\bf{Z}}_M}\left( {{\bf{Z}} - \mu {{\bf{I}}_L}} \right){\bf{BZ}}_G^T,
\end{equation}
where ${{\bf{I}}_L}$ is an $L \times L$ identity matrix, and $\mu$ denotes the scalar complex variable introduced to construct the matrix pencil. Given that the multipath delays are distinct, the poles $z_{nl}$ are distinct. Furthermore, since the pencil parameter satisfies $L \ll \min(M, G)$, the matrices ${\bf{Z}}_M$ and ${\bf{Z}}_G$ are of full column rank $L$, and the diagonal matrix $\bf B$ is of full rank $L$. Consequently, the rank of the matrix pencil ${{\bf{H}}_1} - \mu {{\bf{H}}_0}$ is $L$. When $\mu = z_{nl}$, the $l$-th diagonal element of $\left( {{\bf{Z}} - \mu {{\bf{I}}_L}} \right)$ becomes zero, which reduces the rank of the matrix pencil to $L-1$. Therefore, the signal poles $\{z_{nl}\}_{l=1}^L$ are the generalized eigenvalues of the matrix pencil ${{\bf{H}}_1} - \mu {{\bf{H}}_0}$.

\begin{lemma}\label{lemma1:Eigenvalues}
The non-zero eigenvalues of the product ${ {{\bf{H}}_1 \bf{H}}_0^\dag}$ are identical to those of $\bf Z$, i.e., $z_{nl}$.
\begin{proof}
Please refer to Appendix~A.
\end{proof}
\end{lemma}

\subsubsection{Practical Estimation via SVD in Noisy Environments}
In practical scenarios, the channel observation is affected by the noise component $n_{ik}^{'}$. Consequently, the constructed Hankel matrices ${\bf{\bar{H}}}_0$ and ${\bf{\bar{H}}}_1$ are of full rank $\min(M, G)$ rather than $L$. To mitigate the noise effect and extract the signal subspace, SVD is applied to ${\bf{\bar{H}}}_0$:
\begin{equation}\label{SVD}
{{\bf{\bar H}}_0} = {\bf{U}}\Sigma {{\bf{V}}^H},
\end{equation}
where $\bf U$ and $\bf V$ are unitary matrices containing the left and right singular vectors, respectively, and $\Sigma  = {\rm{diag}}\left( {{\sigma _1},{\sigma _2}, \ldots ,{\sigma _{\min (M,G)}}} \right)$ contains the singular values in descending order. 

As the singular values transform from the signal subspace to the noise subspace, an abrupt drop typically occurs, and the ratio of consecutive singular values reaches its maximum. Therefore, the index corresponding to the sharp drop can be utilized to estimate the number of paths $\hat{L}$, which can be written as:
\begin{equation}\label{estimated number of paths}
\hat L = \arg \mathop {\max }\limits_m \left\{ {\frac{{{\sigma _m}}}{{{\sigma _{m + 1}}}}} \right\},
\end{equation}
where $m \in \{ 1, \ldots ,\min \left( {M,G} \right) - 1\} $.

Based on $\hat{L}$, we extract the signal subspace by truncating the left singular matrix as ${{\bf{U}}_s} = {\bf{U}}\left( {:,1:\hat L} \right)$. The noisy Hankel matrices are then projected onto the robust signal subspace to filter out the noise, which can be expressed as:
\begin{equation}\label{robust signal subspace 0}
{{\bf{H}}_{0s}} = {\bf{U}}_s^H{{\bf{\bar H}}_0},
\end{equation}
and
\begin{equation}\label{robust signal subspace 1}
{{\bf{H}}_{1s}} = {\bf{U}}_s^H{{\bf{\bar H}}_1}.
\end{equation}

Hence, following the mathematical principle derived in {\bf {Lemma}}~\ref{lemma1:Eigenvalues}, the signal poles are obtained by computing the eigenvalues of the projected matrix product:
\begin{equation}\label{signal poles}
{\hat z_{nl}} = {\rm{eig}}\left( {{{\bf{H}}_{1s}} {\bf{H}}_{0s}^\dag } \right).
\end{equation}

\subsubsection{Delay and Amplitude Extraction}
Theoretically, the true poles $z_{nl} = e^{-j2\pi \Delta f \tau_{nl}}$ lie exactly on the unit circle in the complex plane, i.e., $|z_{nl}| = 1$. However, the residual noise may cause the estimated poles to deviate slightly. To further suppress the spurious poles caused by noise, a unit circle filter is applied, which can be written as:
\begin{equation}\label{unit circle filter}
\big| |\hat{z}_{nl}| - 1 \big| \le \epsilon,
\end{equation}
where $\epsilon$ is a small tolerance threshold. For the valid poles that pass this filter, the corresponding multipath delays can be estimated by extracting their phase angles, which can be written as:
\begin{equation}\label{multipath delays}
{\hat \tau _{nl}} =  - \frac{{\arg \left( {{{\hat z}_{nl}}} \right)}}{{2\pi \Delta f}}.
\end{equation}

To robustly identify the LoS path among the valid multipath components, we further estimate the path gain coefficient $\alpha_{nl}$. Let $\bar{\mathbf{h}}_n =[\bar{h}_{n0}, \bar{h}_{n1}, \dots, \bar{h}_{n(K-1)}]^T$ be the channel observation vector, and $\mathbf{\Phi}$ be a $K \times \hat{L}$ matrix where $[\mathbf{\Phi}]_{k,l} = (\hat{z}_{nl})^k$. The complex amplitudes can be computed as:
\begin{equation}\label{complex amplitudes}
    \hat{\boldsymbol{\beta}}_n = \mathbf{\Phi}^\dagger \bar{\mathbf{h}}_n.
\end{equation}
Since $\alpha_{nl}$ represents a real-valued positive attenuation, the magnitude can be estimated by taking the absolute value of the complex amplitude, i.e., $\hat{\alpha}_{nl} = |\hat{\beta}_{nl}|$. The estimated amplitudes of the LoS components can be used to construct the weighting matrix for position estimation.

Finally, based on the shortest estimated delay ${\hat \tau _{n1}}$, the distance of the LoS path is calculated as follows:
\begin{equation}\label{distance of LoS path}
\hat{d}_{n1} = c {\hat \tau _{n1}}.
\end{equation}

\subsubsection{Complexity Analysis}

The computational complexity of the MP-based ranging algorithm is primarily dominated by the full SVD and the pseudo-inverse of matrix. 
To begin with, the phase alignment and coherent averaging over $I$ pilot require $\mathcal{O}(IK)$ operations. Then, performing the full SVD on the $M \times G$ Hankel matrix $\bar{\mathbf{H}}_0$ requires a computational complexity of ${\cal O}\left( {MG\min \left\{ {M,G} \right\}} \right)$~\cite{SVD}. Furthermore, computing the pseudo-inverse of the projected matrix ${\bf{H}}_{0s}$ and the subsequent eigenvalue decomposition of the $\hat{L} \times \hat{L}$ matrix require the computational complexity of ${\cal O}\left( {{{\hat L}^2}G} \right)$ and ${\cal O}\left( \hat{L}^3 \right)$, respectively. Therefore, the computational complexity of the MP-based ranging algorithm scales as ${\cal O}\left( {{\kappa ^3}} \right)$, which can be written as:
\begin{equation}\label{complexity of MP}
{\mathcal{C}_{MP}} = {\cal O}\left( IK + {MG\min \left\{ {M,G} \right\} + {{\hat L}^2}G + {{\hat L}^3}} \right).
\end{equation}

\subsection{Low-Complexity Rank-1 Ranging Algorithm}

While the MP-based ranging algorithm in Section III.A can resolve all multipath components and yield comprehensive channel state information, it comes with a high computational cost. However, for PASS-based indoor positioning, the geometric location of the user is determined by the LoS paths. Although the MP algorithm can accurately resolve the NLoS paths, their contribution to the final position estimation is marginal. Furthermore, in practical mmWave environments, the LoS component typically dominates the received signal power. Therefore, to facilitate practical implementation, we propose a low-complexity Rank-1 ranging algorithm. Instead of jointly estimating all multipath parameters, the simplified method directly isolates the dominant LoS path, thereby avoiding complex matrix inversions while maintaining high robustness against noise. The specific steps are shown in Algorithm \ref{alg:Rank-one ranging}. 

\begin{algorithm}[!ht]
    \renewcommand{\algorithmicrequire}{\textbf{Input:}}
	\renewcommand{\algorithmicensure}{\textbf{Output:}}
	\caption{Low-Complexity Rank-1 Ranging Algorithm}
    \label{alg:Rank-one ranging}
    \begin{algorithmic}[1] 
        \REQUIRE  Averaged channel observation $\bar{h}_{nk}$, waveguide parameters, the number of subcarriers $K$, subcarrier spacing $\Delta f$. 
	    \ENSURE The estimated LoS distance $\hat {d}_{n1}$. 
        \STATE \textbf{Hankel matrix construction:} Select the pencil parameter $M$ and construct the Hankel matrix $\mathbf{H}_n \in \mathbb{C}^{(M+1) \times G}$ based on $\bar{h}_{nk}$.
        \STATE \textbf{Rank-1 approximation:} Perform truncated SVD on $\mathbf{H}_n$ to extract the largest singular value ${\sigma _{R,1}}$ and corresponding singular vectors ${{\bf{u}}_{R,1}}{\bf{v}}_{R,1}$. Then, reconstruct the pure LoS matrix ${{\bf{\tilde H}}_n} = {\sigma _{R,1}}{{\bf{u}}_{R,1}}{\bf{v}}_{R,1}^H$.
        \STATE \textbf{Submatrix partitioning:} Partition ${{\bf{\tilde H}}_n}$ to $\tilde{\mathbf{H}}_{n,A}$ and $\tilde{\mathbf{H}}_{n,B}$ by extracting columns 1 to $G-1$ and 2 to $G$, respectively.        
        \STATE \textbf{Pole estimation:} Compute the LoS signal pole $\hat{z}_{n1}$ via LS.       
        \STATE \textbf{Delay and distance calculation:} The delay and distance are calculated as in~\eqref{unit circle filter},~\eqref{multipath delays} and~\eqref{distance of LoS path}.
        \RETURN The estimated LoS distance $\hat{d}_{n1}$
    \end{algorithmic}
\end{algorithm}

\subsubsection{Hankel Matrix Construction and Rank-1 Approximation}

Based on the coherent averaged channel $\bar{h}_{nk}$ obtained in \eqref{eq:averaged_channel}, we construct a Hankel matrix $\mathbf{H}_n \in \mathbb{C}^{(M+1) \times G}$ as follows:
\begin{equation}\label{Hankel matrix n}
    \mathbf{H}_n = 
    \begin{bmatrix}
        \bar{h}_{n0} & \bar{h}_{n1} & \cdots & \bar{h}_{n(G-1)} \\
        \bar{h}_{n1} & \bar{h}_{n2} & \cdots & \bar{h}_{nG} \\
        \vdots & \vdots & \ddots & \vdots \\
        \bar{h}_{nM} & \bar{h}_{n(M+1)} & \cdots & \bar{h}_{n(K-1)}
    \end{bmatrix}.
\end{equation}

In the MP method, the effective rank of the Hankel matrix $\mathbf{H}_n$ is determined by the number of multipath components $L$. However, since the power of the LoS path is significantly higher than that of the NLoS paths, the principal energy of the Hankel matrix $\mathbf{H}_n$ is captured by its largest singular value. To extract the LoS component with low complexity, we approximate $\mathbf H_n$ by its rank-one truncated SVD, which can be written as:
\begin{equation}\label{truncated SVD}
\widetilde{\mathbf H}_n
= \sigma_{R,1}\mathbf u_{R,1}\mathbf v_{R,1}^H,
\end{equation}
where $\sigma_{R,1}$ is the largest singular value of $\mathbf H_n$, and $\mathbf u_{R,1}$ and $\mathbf v_{R,1}$ are the associated left and right singular vectors, respectively.

According to the Eckart-Young-Mirsky theorem, $\tilde{\mathbf{H}}_n$ is the optimal Rank-1 approximation of $\mathbf{H}_n$ in the sense of the Frobenius norm~\cite{matrix_analysis}. The Rank-1 approximation eliminates the subspace spanned by NLoS paths and noise. 

\subsubsection{Pole Extraction via Least Squares}

Let $\tilde{h}_{p,q}$ denote the element in the $p$-th row and $q$-th column of the Rank-1 matrix $\tilde{\mathbf{H}}_n$. We partition $\tilde{\mathbf{H}}_n$ into two overlapping submatrices:
\begin{equation}\label{submatrices 1}
    \tilde{\mathbf{H}}_{n,A} = 
    \begin{bmatrix}
        \tilde{h}_{0,0} & \tilde{h}_{0,1} & \cdots & \tilde{h}_{0,G-2} \\
        \tilde{h}_{1,0} & \tilde{h}_{1,1} & \cdots & \tilde{h}_{1,G-2} \\
        \vdots & \vdots & \ddots & \vdots \\
        \tilde{h}_{M,0} & \tilde{h}_{M,1} & \cdots & \tilde{h}_{M,G-2}
    \end{bmatrix},
\end{equation}

\begin{equation}\label{submatrices 1}
    \tilde{\mathbf{H}}_{n,B} = 
    \begin{bmatrix}
        \tilde{h}_{0,1} & \tilde{h}_{0,2} & \cdots & \tilde{h}_{0,G-1} \\
        \tilde{h}_{1,1} & \tilde{h}_{1,2} & \cdots & \tilde{h}_{1,G-1} \\
        \vdots & \vdots & \ddots & \vdots \\
        \tilde{h}_{M,1} & \tilde{h}_{M,2} & \cdots & \tilde{h}_{M,G-1}
    \end{bmatrix}.
\end{equation}

Since $\tilde{\mathbf{H}}_n$ is a Rank-1 matrix containing only the LoS component, the shift-invariance relationship simplifies to a scalar multiplication:
\begin{equation}\label{shift-invariance relationship}
    \tilde{\mathbf{H}}_{n,B} = z_{n1} \tilde{\mathbf{H}}_{n,A}.
\end{equation}

To obtain an accurate pole estimate, we formulate an LS problem by minimizing the residual of Frobenius norm between two shifted submatrices, which is given by:
\begin{equation}\label{accurate pole estimate}
    J(z) = ||\tilde{\mathbf{H}}_{n,B} - z \tilde{\mathbf{H}}_{n,A}||_F^2.
\end{equation}

By using the property of the Frobenius norm, the function $J(z)$ can be expanded as:
\begin{equation}\label{expanded function}
\begin{aligned}
J\left( z \right) &= {\rm{Tr}}\left( {{{\left( {{{\widetilde {\bf{H}}}_{n,B}} - z{{\widetilde {\bf{H}}}_{n,A}}} \right)}^H}\left( {{{\widetilde {\bf{H}}}_{n,B}} - z{{\widetilde {\bf{H}}}_{n,A}}} \right)} \right)\\
 &= {\rm{Tr}}\left( {\widetilde {\bf{H}}_{n,B}^H{{\widetilde {\bf{H}}}_{n,B}}} \right) - z{\rm{Tr}}\left( {\widetilde {\bf{H}}_{n,B}^H{{\widetilde {\bf{H}}}_{n,A}}} \right)\\
 &- {z^*}{\rm{Tr}}\left( {\widetilde {\bf{H}}_{n,A}^H{{\widetilde {\bf{H}}}_{n,B}}} \right) + |z{|^2}{\rm{Tr}}\left( {\widetilde {\bf{H}}_{n,A}^H{{\widetilde {\bf{H}}}_{n,A}}} \right).
\end{aligned}
\end{equation}

According to Wirtinger calculus for complex optimization, we take the partial derivative of $J(z)$ with respect to the conjugate variable $z^*$ and set it to zero:
\begin{equation}\label{partial derivative}
\frac{{\partial J\left( z \right)}}{{\partial {z^*}}} =  - {\rm{Tr}}\left( {\widetilde {\bf{H}}_{n,A}^H{{\widetilde {\bf{H}}}_{n,B}}} \right) + z{\rm{Tr}}\left( {\widetilde {\bf{H}}_{n,A}^H{{\widetilde {\bf{H}}}_{n,A}}} \right) = 0.
\end{equation}

By solving \eqref{partial derivative}, the closed-form solution for the LoS pole can be written as:
\begin{equation}\label{closed-form solution}
{\hat z_{n1}} = \frac{{{\rm{Tr}}\left( {\widetilde {\bf{H}}_{n,A}^H{{\widetilde {\bf{H}}}_{n,B}}} \right)}}{{{\rm{Tr}}\left( {\widetilde {\bf{H}}_{n,A}^H{{\widetilde {\bf{H}}}_{n,A}}} \right)}}.
\end{equation}

\subsubsection{Delay and Amplitude Estimation}

After estimating the complex pole $\hat{z}_{n1}$, the delay and distance of the LoS path can be calculated as in~\eqref{unit circle filter},~\eqref{multipath delays} and~\eqref{distance of LoS path}. Furthermore, to estimate the physical amplitude of the LoS path, we construct the vector based on the estimated pole $\hat{z}_{n1}$, which can be written as:
\begin{equation}\label{steering vector}
    \mathbf{a}_n = \left[1, \hat{z}_{n1}, \hat{z}_{n1}^2, \dots, \hat{z}_{n1}^{K-1}\right]^T.
\end{equation}

Since the Rank-1 approximation effectively isolates the LoS component, the original coherent averaged channel vector can be approximated as $\bar{\mathbf{h}}_n \approx \beta_{n1} \mathbf{a}_n$. Therefore, the complex amplitude $\beta_{n1}$ can be estimated by projecting $\bar{\mathbf{h}}_n$ onto the steering vector $\mathbf{a}_n$ via the LS criterion:
\begin{equation}\label{Rank-1 complex amplitude}
    \hat{\beta}_{n1} = \frac{\mathbf{a}_n^H \bar{\mathbf{h}}_n}{\mathbf{a}_n^H \mathbf{a}_n}.
\end{equation}
Similarly, we can obtain $\hat{\alpha}_{n1} = |\hat{\beta}_{n1}|$.

\subsubsection{Complexity Analysis}

To start with, the computational cost of phase alignment and coherent averaging is $\mathcal{O}(IK)$. For the Rank-1 approximation, the truncated SVD only requires a complexity of $\mathcal{O}(MG)$. Then, the trace operations for the LS pole estimation merely involve element-wise multiplications and summations, which require the complexity of ${\cal O}\left( {\left( {M + 1} \right)\left( {G - 1} \right)} \right)$. Therefore, the computational complexity of Algorithm~\ref{alg:Rank-one ranging} scales as ${\cal O}\left( {{\kappa ^2}} \right)$, which can be written as:
\begin{equation}\label{complexity of rank one}
\begin{aligned}
{\mathcal{C}_R} &= {\cal O}\left( {IK + MG + \left( {M + 1} \right)\left( {G - 1} \right)} \right)\\
 &= {\cal O}\left( {IK + 2MG + G - M - 1} \right)\\
 &= {\cal O}\left( {IK + MG} \right).
\end{aligned}
\end{equation} 

The proposed Rank-1 ranging algorithm significantly reduces the computational complexity by avoiding full matrix decompositions and inversions, which is critical for the real-time indoor positioning systems.

\subsection{Two-Stage WNLS Positioning Algorithm}

After extracting the LoS distances $\{\hat{d}_{n1}\}_{n=1}^4$ from the multipath environment by using the proposed ranging algorithms, the 3D position of the user can then be estimated. Algorithm~\ref{alg:positioning} illustrates the process of position calculation. 

\begin{algorithm}[t]
\caption{Two-Stage WNLS Positioning Algorithm}
\label{alg:positioning}
\begin{algorithmic}[1] 
\renewcommand{\algorithmicrequire}{\textbf{Input:}}
\renewcommand{\algorithmicensure}{\textbf{Output:}}
\REQUIRE The PAs positions $\{(x_n, y_n, z_n)\}_{n=1}^4$, estimated LoS distances $\{\hat{d}_{n1}\}_{n=1}^4$, weight matrix $\mathbf{W}$, tolerance $\varepsilon$, maximum iterations $t_{\max}$.
\ENSURE Estimated user position $\hat{\mathbf{X}}$.

\textbf{Stage 1: Initial Estimation}
\STATE Construct the distance equations, and subtract the first equation from the remaining equations to obtain the difference equation set.
\STATE Construct matrix $\mathbf{A}$ and vector $\mathbf{E}$ according to \eqref{A} and \eqref{E}.
\STATE Compute the initial position estimate $\hat{\mathbf{X}}_0$.

\textbf{Stage 2: WNLS Refinement}
\STATE Initialize $\mathbf{X}^{(0)} = \hat{\mathbf{X}}_0$, and iteration index $t = 0$.
\REPEAT
    \item[] Compute residual vector $\mathbf{r}(\mathbf{X}^{(t)})$ by \eqref{residual function}.
    \item[] Compute Jacobian matrix $\mathbf{J}(\mathbf{X}^{(t)})$ by \eqref{Jacobian matrix}.
    \item[] Calculate the Damped Newton direction $\mathbf{p}^{(t)}$ by \eqref{Newton direction}.
    \item[] Update the position estimate: $\mathbf{X}^{(t+1)} = \mathbf{X}^{(t)} + \alpha^{(t)}\mathbf{p}^{(t)}$.
    \item[] Update iteration index: $t \leftarrow t + 1$.
\UNTIL{$\|\mathbf{X}^{(t)} - \mathbf{X}^{(t-1)}\|_2 \le \varepsilon$ or $t = t_{\max}$}

\STATE \textbf{Return} Estimated position $\hat{\mathbf{X}} = \mathbf{X}^{(t)}$.
\end{algorithmic}
\end{algorithm}

To begin with, the distance equation is given by:
\begin{equation}\label{diatance equation}
\left\{ {\begin{array}{*{20}{c}}
{{{\left( {{x_1} - {x_u}} \right)}^2} + {{\left( {{y_1} - {y_u}} \right)}^2} + {{\left( {{z_1} - {z_u}} \right)}^2} = \hat d_{11}^2,}\\
{{{\left( {{x_2} - {x_u}} \right)}^2} + {{\left( {{y_2} - {y_u}} \right)}^2} + {{\left( {{z_2} - {z_u}} \right)}^2} = \hat d_{21}^2,}\\
{{{\left( {{x_3} - {x_u}} \right)}^2} + {{\left( {{y_3} - {y_u}} \right)}^2} + {{\left( {{z_3} - {z_u}} \right)}^2} = \hat d_{31}^2,}\\
{{{\left( {{x_4} - {x_u}} \right)}^2} + {{\left( {{y_4} - {y_u}} \right)}^2} + {{\left( {{z_4} - {z_u}} \right)}^2} = \hat d_{41}^2.}
\end{array}} \right.
\end{equation}

By subtracting the first equation from the remaining equations, the quadratic term of the user position is eliminated, and we can obtain:
\begin{equation}\label{linear distance equation}
\left\{ {\begin{array}{*{20}{c}}
{2{x_u}\left( {{x_2} - {x_1}} \right) + 2{y_u}\left( {{y_2} - {y_1}} \right) + 2{z_u}\left( {{z_2} - {z_1}} \right) = {e_1},}\\
{2{x_u}\left( {{x_3} - {x_1}} \right) + 2{y_u}\left( {{y_3} - {y_1}} \right) + 2{z_u}\left( {{z_3} - {z_1}} \right) = {e_2},}\\
{2{x_u}\left( {{x_4} - {x_1}} \right) + 2{y_u}\left( {{y_4} - {y_1}} \right) + 2{z_u}\left( {{z_4} - {z_1}} \right) = {e_3},}
\end{array}} \right.
\end{equation}
where 
\begin{equation}\label{ei}
\begin{aligned}
{e_{i - 1}} &= \hat d_{11}^2 - \hat d_{i1}^2 + \left( {x_i^2 - x_1^2} \right) + \left( {y_i^2 - y_1^2} \right) + \left( {z_i^2 - z_1^2} \right),\\
i &= 2,3,4.
\end{aligned}
\end{equation}

Then, \eqref{linear distance equation} can be compactly written in the matrix form as:
\begin{equation}\label{matrix distance equation}
{\bf{AX}} = {\bf{E}},
\end{equation}
where
\begin{equation}\label{A}
{\bf{A}} = 2\left[ {\begin{array}{*{20}{c}}
{\left( {{x_2} - {x_1}} \right)}&{\left( {{y_2} - {y_1}} \right)}&{\left( {{z_2} - {z_1}} \right)}\\
{\left( {{x_3} - {x_1}} \right)}&{\left( {{y_3} - {y_1}} \right)}&{\left( {{z_3} - {z_1}} \right)}\\
{\left( {{x_4} - {x_1}} \right)}&{\left( {{y_4} - {y_1}} \right)}&{\left( {{z_4} - {z_1}} \right)}
\end{array}} \right],
\end{equation}
\begin{equation}\label{X}
{\bf{X}} = {\left[ {\begin{array}{*{20}{c}}
{{x_u}}&{{y_u}}&{{z_u}}
\end{array}} \right]^T},\\
\end{equation}
\begin{equation}\label{E}
{\bf{E}} = {\left[ {\begin{array}{*{20}{c}}
{{e_1}}&{{e_2}}&{{e_3}}
\end{array}} \right]^T}.
\end{equation}

Then, the position estimation ${{\bf{\hat X}}_0}$ is given by:
\begin{equation}\label{position solution}
{{\bf{\hat X}}_0} = {\left( {{{\bf{A}}^T}{\bf{A}}} \right)^{ - 1}}{{\bf{A}}^T}{\bf{E}}.
\end{equation}

Since the estimated distances $\{\hat{d}_{n1}\}_{n=1}^4$ are generally affected by ranging errors, the linear solution in \eqref{position solution} may suffer from the performance degradation in practical multipath environments. To further improve the positioning accuracy, we refine the initial estimate ${{\bf{\hat X}}_0}$ by introducing the following residual function:
\begin{equation}\label{residual function}
\begin{aligned}
r_n(\mathbf X)=\sqrt{(x_n-x_u)^2+(y_n-y_u)^2+(z_n-z_u)^2}-\hat d_{n1}.
\end{aligned}
\end{equation}

Therefore, the accurate position estimate is obtained by solving a weighted nonlinear least-squares (WNLS) problem, which can be written as:
\begin{equation}\label{weighted nonlinear least-squares}
\hat{\mathbf X}
=\arg\min_{\mathbf X}\sum_{n=1}^{4} w_n\, r_n^2(\mathbf X),
\end{equation}
where $w_n$ denotes the weighting coefficient of the $n$-th waveguide. Since a stronger LoS component generally indicates a more reliable ranging estimation, $w_n$ can be chosen according to the estimated amplitude of the corresponding LoS component in \eqref{complex amplitudes} and \eqref{Rank-1 complex amplitude}.

The NLS problem can be rewritten as:
\begin{equation}\label{rewritten NLS problem}
f(\mathbf X)=\frac{1}{2}\mathbf r^T(\mathbf X)\mathbf W \mathbf r(\mathbf X),
\end{equation}
where $t$ represents the number of iterations, and
\begin{equation}\label{rX}
{\bf{r}}({\bf{X}}) = {\left[ {\begin{array}{*{20}{c}}
{{r_1}({\bf{X}})}&{{r_2}({\bf{X}})}&{{r_3}({\bf{X}})}&{{r_4}({\bf{X}})}
\end{array}} \right]^T},
\end{equation}
\begin{equation}\label{W}
{\bf{W}} = {\rm{diag}}\left( {{w_1},{w_2},{w_3},{w_4}} \right).
\end{equation}

Then, the Newton direction at the $t$-th iteration is given by~\cite{Newton}:
\begin{equation}\label{Newton direction}
\mathbf p^{(t)}
=
-
\left(\mathbf J^T\mathbf W\mathbf J+\lambda^{(t)}\mathbf I\right)^{-1}
\mathbf J^T\mathbf W\mathbf r\!\left(\mathbf X^{(t)}\right),
\end{equation}
where $\lambda^{(t)}$ is the damping factor at the $t$-th iteration, and $\mathbf J$ denotes the Jacobian matrix of $\mathbf r(\mathbf X)$ with respect to $\mathbf X$, which can be written as:
\begin{equation}\label{Jacobian matrix}
\mathbf J(\mathbf X)=
\frac{\partial \mathbf r(\mathbf X)}{\partial \mathbf X}
=
\begin{bmatrix}
\dfrac{x_u-x_1}{\hat d_{11}(\mathbf X)} & \dfrac{y_u-y_1}{\hat d_{11}(\mathbf X)} & \dfrac{z_u-z_1}{\hat d_{11}(\mathbf X)}\\[1.2ex]
\dfrac{x_u-x_2}{\hat d_{21}(\mathbf X)} & \dfrac{y_u-y_2}{\hat d_{21}(\mathbf X)} & \dfrac{z_u-z_2}{\hat d_{21}(\mathbf X)}\\[1.2ex]
\dfrac{x_u-x_3}{\hat d_{31}(\mathbf X)} & \dfrac{y_u-y_3}{\hat d_{31}(\mathbf X)} & \dfrac{z_u-z_3}{\hat d_{31}(\mathbf X)}\\[1.2ex]
\dfrac{x_u-x_4}{\hat d_{41}(\mathbf X)} & \dfrac{y_u-y_4}{\hat d_{41}(\mathbf X)} & \dfrac{z_u-z_4}{\hat d_{41}(\mathbf X)}
\end{bmatrix}.
\end{equation}

The estimated position is updated as:
\begin{equation}\label{accurate estimated position}
\mathbf X^{(t+1)}=\mathbf X^{(t)}+\alpha^{(t)}\mathbf p^{(t)},
\end{equation}
where \(\alpha^{(t)}\in(0,1]\) denotes the step size. In this paper, a unit step size is adopted, i.e., \(\alpha^{(t)}=1\).

The iteration is terminated when the change between two consecutive estimates satisfies $\left\|\mathbf X^{(t+1)}-\mathbf X^{(t)}\right\|_2 \le \varepsilon$, where \(\varepsilon\) is a prescribed tolerance. In addition, a maximum number of iterations $t_{\max}$ is imposed to guarantee termination. Therefore, the final position estimate is given by:
\begin{equation}\label{final position estimate}
\hat{\mathbf X}=\mathbf X^{(t_{m})},
\end{equation}
where $t_{m}$ denotes the iteration index at convergence or the maximum iteration index.

\section{Performance Analysis}

To comprehensively evaluate the performance of the proposed PASS-based indoor positioning system, we establish a theoretical framework in this section. We first derive the theoretical ranging variance of the proposed ranging algorithms. Subsequently, we map the ranging errors to the 3D PEB through error propagation analysis.

\subsection{Performance Analysis of MP-based Ranging Algorithm}

We first investigate the ranging performance of the MP-based algorithm. After coherent averaging over $I$ pilot symbols, the equivalent channel observation at the $n$-th PA and the $k$-th subcarrier can be approximated as:
\begin{equation}\label{equivalent channel}
{\bar h_{nk}} \approx {\beta _{n1}}z_{n1}^k + {w_{nk}},
\end{equation}
where $w_{nk}$ is the zero-mean complex Gaussian noise with variance $\sigma_w^2$, and $\sigma_w^2=\frac{\sigma^2}{IP|h_{w,n}|^2}$.

According to the first-order perturbation analysis of the matrix pencil method, the pole estimation variance in the multipath environment is coupled with the path amplitudes, phases and pole separations, lacking a simple closed-form expression, which can be evaluated numerically using the general perturbation formulas in~\cite{matrix_pencil}. To obtain tractable physical insights, we consider a LoS-dominant scenario under the high SNR assumption. Thus, the asymptotic variance of the LoS phase estimate is given by \cite{matrix_pencil}:
\begin{equation}\label{var_mp}
    \mathrm{Var}\!\left(\arg \left( \hat{z}_{n1}\right)\right)
    \approx
    \frac{{\sigma _w^2}}{{|{\beta _{n1}}{|^2}}}
    \mathcal{F}_{\mathrm{MP}}(M,K),    
\end{equation}
where
\begin{equation}\label{fmp}
\mathcal{F}_{\mathrm{MP}}(M,K)
=
\begin{cases}
\dfrac{1}{\left( K-M \right)^2 M}, & M\leq \dfrac{K}{2}, \\[2ex]
\dfrac{1}{\left( K-M \right) M^2}, & M\geq \dfrac{K}{2}.
\end{cases}
\end{equation}

\begin{theorem}\label{Theorem1:variance of alg1}
\emph{The ranging variance of the MP algorithm is approximated by:}
\begin{equation} \label{var_d_mp}
{V_{MP,n}} \approx {\left( {\frac{c}{{2\pi \Delta f}}} \right)^2}\frac{{\sigma _w^2}}{{|{\beta _{n1}}{|^2}}}{{\cal F}_{{\rm{MP}}}}\left( {M,K} \right).
\end{equation}
\begin{proof}
Since the estimated distance is linearly mapped from the phase as ${\hat d_{n1}} =  - \frac{c}{{2\pi \Delta f}}\arg \left( {{{\hat z}_{nl}}} \right)$, the LoS phase-estimation variance in \eqref{var_mp} can be mapped to the corresponding LoS distance-estimation variance in \eqref{var_d_mp}. The proof is complete.
\end{proof}
\end{theorem}

\subsection{Performance Analysis of the Rank-1 Ranging Algorithm}

In this subsection, we derive the closed-form variance of the distance estimate $\hat{d}_{n1}$ obtained by Algorithm \ref{alg:Rank-one ranging}. The first-order matrix perturbation theory indicates that under the high-SNR assumption, the truncated SVD suppresses the noise components outside the dominant signal subspace, such that the truncated rank-one matrix can be viewed as the noise-free signal matrix plus an equivalent first-order perturbation within the signal subspace. Thus, the partitioned submatrices $\tilde{\mathbf{H}}_{n,A}$ and $\tilde{\mathbf{H}}_{n,B}$ in Algorithm \ref{alg:Rank-one ranging} can be respectively modeled as:
\begin{equation}\label{perturbed submatrices 1}
{\tilde{\mathbf{H}}}_{n,A} = {\mathbf{H}}_{n,A}^{o} + {\mathbf{N}}_{n,A},
\end{equation}
and
\begin{equation}\label{perturbed submatrices 2}
{\tilde{\mathbf{H}}}_{n,B} = {\mathbf{H}}_{n,B}^{o} + {\mathbf{N}}_{n,B},
\end{equation}
where $\mathbf{H}_{n,A}^{\text{o}}$ and $\mathbf{H}_{n,B}^{\text{o}}$ are the noise-free Rank-1 matrices, which strictly satisfy the shift-invariance property $\mathbf{H}_{n,B}^{\text{o}} = z_{n1} \mathbf{H}_{n,A}^{\text{o}}$. $\mathbf{N}_{n,A}, \mathbf{N}_{n,B}$ are the corresponding noise perturbation matrices.

By substituting \eqref{perturbed submatrices 1} and \eqref{perturbed submatrices 2} into \eqref{closed-form solution}, and neglecting the second-order noise terms $\mathbf{N}_{n,A}^H \mathbf{N}_{n,B}$, the estimated pole can be expanded as:
\begin{equation} \label{pole_expand}
\begin{aligned}
{{\hat z}_{n1}} &\approx \frac{{{\mathop{\rm Tr}\nolimits} \left( {{{({\bf{H}}_{n,A}^{\rm{o}})}^H}{\bf{H}}_{n,B}^{\rm{o}} + {{({\bf{H}}_{n,A}^{\rm{o}})}^H}{{\bf{N}}_{n,B}} + {\bf{N}}_{n,A}^H{\bf{H}}_{n,B}^{\rm{o}}} \right)}}{{{\mathop{\rm Tr}\nolimits} \left( {{{({\bf{H}}_{n,A}^{\rm{o}})}^H}{\bf{H}}_{n,A}^{\rm{o}} + {{({\bf{H}}_{n,A}^{\rm{o}})}^H}{{\bf{N}}_{n,A}} + {\bf{N}}_{n,A}^H{\bf{H}}_{n,A}^{\rm{o}}} \right)}}.
\end{aligned}
\end{equation}

\begin{lemma}\label{lemma2:delta}
The pole error of Rank-1 ranging algorithm is given by:
\begin{equation} \label{eq:pole_error_linear}
    \Delta z_{n1} \approx \frac{\mathrm{Tr}\big((\mathbf{H}_{n,A}^{\text{o}})^H(\mathbf{N}_{n,B} - z_{n1}\mathbf{N}_{n,A})\big)}{\|\mathbf{H}_{n,A}^{\text{o}}\|_F^2}.
\end{equation}
\begin{proof}
Please refer to Appendix~B.
\end{proof}
\end{lemma}

\begin{theorem}\label{Theorem2:variance of alg2}
\emph{For the $n$-th PA, the variance of the LoS distance estimated by the Rank-1 ranging algorithm is given by:}
\begin{equation} \label{var_d_rank1}
{V_{R1,n}} = {\left( {\frac{c}{{2\pi \Delta f}}} \right)^2}\frac{{\min (M + 1,G - 1)}}{{{{(M + 1)}^2}{{(G - 1)}^2}}}\frac{{\sigma _w^2}}{{|{\beta _{n1}}{|^2}}}.
\end{equation}
\begin{proof}
Please refer to Appendix C.
\end{proof}
\end{theorem}

\begin{remark}\label{remark1:subcarrier}
According to \cite{matrix_pencil}, the pencil parameter $M$ is typically chosen within the range $\frac{K}{3} \le M \le \frac{{2K}}{3}$. Therefore, the ranging performance of the MP-based and Rank-1-based algorithms improves as the number of subcarriers increases.
\end{remark}

\subsection{PEB for WNLS Positioning Algorithm}

Based on the ranging variance derived above, we can analyze PEB for the proposed algorithm. For a given ranging algorithm $\mathcal A\in\{\mathrm{MP},\mathrm{R1}\}$, the estimated LoS distances can be modeled as:
\begin{equation}\label{eq:range_measurement_model}
{\widehat {\bf{d}}_{\cal A}} = {\bf{d}}\left( {\bf{X}} \right) + {{\bf{e}}_{\cal A}},
\end{equation}
where
\begin{equation}
{\widehat {\bf{d}}_{\cal A}} = {[\begin{array}{*{20}{c}}
{\hat d_{11}^{\cal A}}&{\hat d_{21}^{\cal A}}&{\hat d_{31}^{\cal A}}&{\hat d_{41}^{\cal A}}
\end{array}]^T},
\end{equation}
\begin{equation}
    {\bf{d}}\left( {\bf{X}} \right) = {[\begin{array}{*{20}{c}}
{{d_{11}}}&{{d_{21}}}&{{d_{31}}}&{{d_{41}}}
\end{array}]^T},
\end{equation}
and $\mathbf e_{\mathcal A}$ denotes the ranging error vector. Under the high-SNR approximation, the ranging errors are modeled as zero-mean Gaussian variables, which can be written as:
\begin{equation}\label{ranging error vector}
    \mathbf e_{\mathcal A}
    \sim
    \mathcal N(\mathbf 0,\mathbf C_{\mathcal A}),
\end{equation}
where
\begin{equation}
    \mathbf C_{\mathcal A}
    =
    \mathrm{diag}
    \left(
    V_{\mathcal A,1},
    V_{\mathcal A,2},
    V_{\mathcal A,3},
    V_{\mathcal A,4}
    \right).
    \label{eq:range_error_covariance}
\end{equation}

Then, the log-likelihood function of $\hat{\mathbf d}_{\mathcal A}$ is given by:
\begin{equation}\label{log-likelihood function}
\begin{aligned}
&\ln p({\widehat {\bf{d}}_{\cal A}};{\bf{X}}) = {\rm{const}}\\
& - \frac{1}{2}{\left( {{{\widehat {\bf{d}}}_{\cal A}} - {\bf{d}}({\bf{X}})} \right)^T}{\bf{C}}_{\cal A}^{ - 1}\left( {{{\widehat {\bf{d}}}_{\cal A}} - {\bf{d}}({\bf{X}})} \right).
\end{aligned}
\end{equation}
Therefore, the FIM for the user position can be written as~\cite{FIM}:
\begin{equation}\label{eq:fim_general}
    \mathbf I_{\mathcal A}(\mathbf X)
    =
    \left(
    \frac{\partial \mathbf d(\mathbf X)}{\partial \mathbf X}
    \right)^T
    \mathbf C_{\mathcal A}^{-1}
    \left(
    \frac{\partial \mathbf d(\mathbf X)}{\partial \mathbf X}
    \right).   
\end{equation}
Since
\begin{equation}
    \frac{\partial d_{n1}(\mathbf X)}{\partial \mathbf X}
    =
    \left[
    \frac{x_u-x_n}{d_{n1}},
    \frac{y_u-y_n}{d_{n1}},
    \frac{z_u-z_n}{d_{n1}}
    \right],
\end{equation}
the FIM can be explicitly written as:
\begin{equation}\label{FIM}
\mathbf{I}_{\mathcal{A}}(\mathbf{X}) = \sum_{n=1}^{4} \frac{1}{V_{\mathcal{A},n} d_{n1}^2} \mathbf{v}_n \mathbf{v}_n^T,
\end{equation}
where $\mathbf{v}_n = [x_u - x_n, y_u - y_n, z_u - z_n]^T$.

Since the covariance matrix of any unbiased position estimator satisfies ${\rm{Cov}}(\widehat {\bf{X}}) \ge {\bf{I}}_{\cal A}^{ - 1}({\bf{X}})$, the PEB is given by:
\begin{equation}\label{eq:peb}
    \mathrm{PEB}_{\mathcal A}
    =
    \sqrt{
    \mathrm{tr}
    \left(
    \mathbf I_{\mathcal A}^{-1}(\mathbf X)
    \right)
    }.
\end{equation}
For the MP-based and Rank-1 ranging algorithms, the corresponding bounds are obtained by substituting $\mathbf C_{\mathrm{MP}}$ and $\mathbf C_{\mathrm{R1}}$ into \eqref{eq:fim_general}, respectively.

\section{Simulation Results}

In this section, Monte Carlo simulations are conducted to evaluate both the effectiveness and performance of the proposed ranging and positioning algorithms. For each simulation setting, we generate 200 random multipath environments, with 1000 independent ranging and positioning simulations performed in each environment. The room dimensions are defined as 6 m × 10 m × 3 m. The number of PAs is set to $N$ = 4. The pencil parameter satisfies $M = K/2$. The dielectric properties are specified by a relative permittivity of ${\varepsilon _r = 1.03}$ and a loss tangent of $\tan \delta = 0.0001$~\cite{Microwave_engineering}. The carrier frequency is fixed at $f_c$ = 18 GHz. Following typical 5G communication settings, we adopt a signal bandwidth of 100 MHz and a transmit power of 0.1 W.

\begin{figure}[t!]
\centering
\includegraphics[width =3.5in]{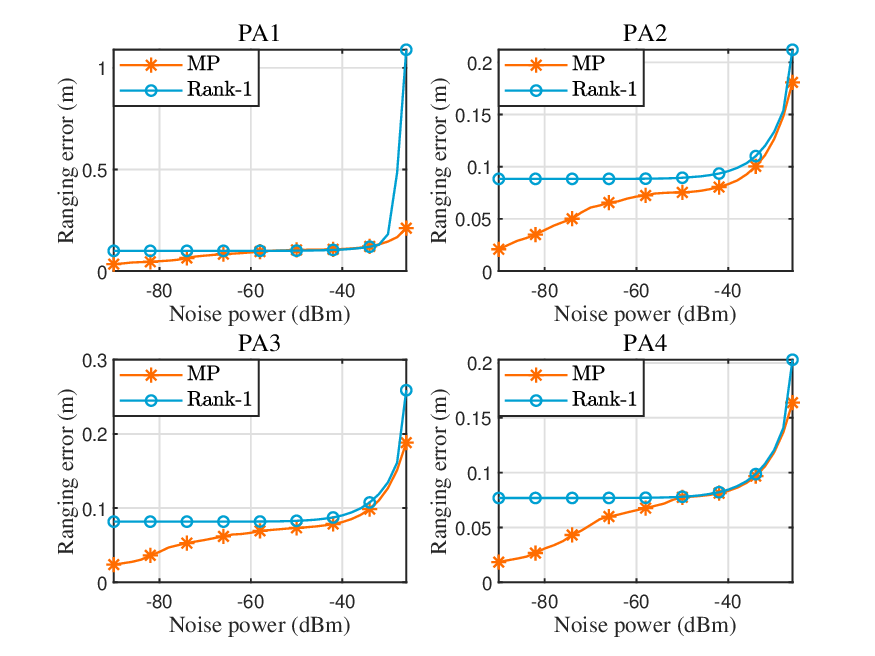}
\caption{The ranging errors of MP and Rank-1 algorithms, where the subcarrier number is set to $K$ = 256.}
\label{ranging error}
\end{figure}

\begin{figure}[t!]
\centering
\includegraphics[width =3.5in]{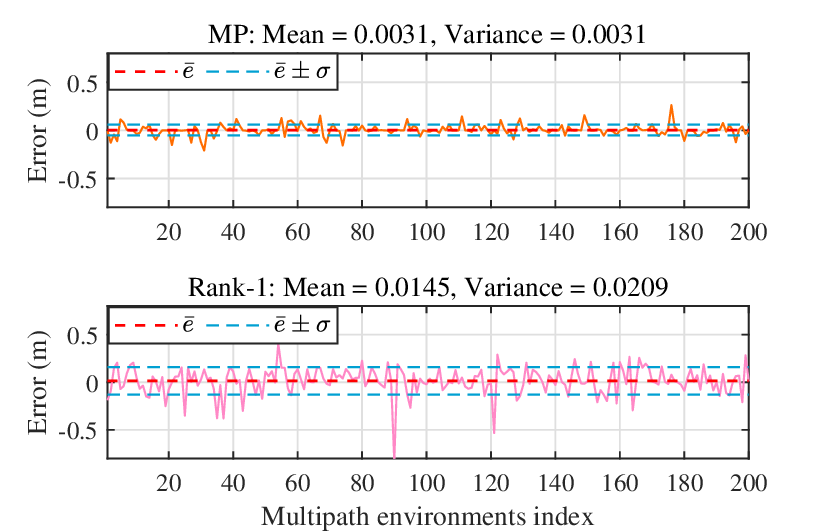}
\caption{Mean and variance of ranging error. The noise power is set to -90 dBm, and the number of subcarriers is 256.}
\label{Robust}
\end{figure}

\emph{1) Ranging accuracy and robustness of MP and Rank-1 algorithms:} Fig.~\ref{ranging error} compares the ranging errors of the MP and Rank-1 algorithms for four PAs under different noise powers. When the noise power is low, the MP algorithm achieves higher ranging accuracy than the Rank-1 algorithm. As the noise power decreases, the ranging error of MP algorithm reduces monotonically, while the Rank-1 algorithm exhibits a clear error floor. The MP algorithm separates the NLoS components from the LoS path, thereby enabling nearly unbiased LoS estimation. By contrast, the Rank-1 algorithm approximates the received signal using a one-dimensional signal subspace, leaving residual NLoS components as structured interference. Therefore, when the noise power is low, this deterministic multipath interference dominates the ranging error of Rank-1 algorithm, leading to the observed error floor. As the noise power increases, the performance gap between the two algorithms gradually narrows, since random noise becomes the dominant error source. Once the noise power exceeds a certain threshold, both algorithms experience rapid performance degradation. This is because excessive noise causes severe perturbations, making it difficult to distinguish the signal subspace from the noise subspace. In addition, since the user-to-PA distances and the relative strengths of the NLoS components vary across links, the lower bounds of the ranging errors differ across the four PAs.

To further examine the robustness of two ranging algorithms against random multipath variations, Fig.~\ref{Robust} depicts the ranging errors over 200 independently generated multipath environments at a noise power of $-90$ dBm, taking $n$ = 1 as an example. It can be observed that the MP algorithm achieves smaller mean error and variance than that of the Rank-1 algorithm, which indicates that the MP algorithm is less sensitive to the changes in the multipath environments. In contrast, the Rank-1 algorithm exhibits larger error fluctuations and occasional severe deviations. This is because the one-dimensional subspace approximation cannot sufficiently suppress the residual NLoS components, whose amplitudes, phases, and delays vary across environments. Therefore, Fig.~~\ref{Robust} further confirms that the MP algorithm provides higher ranging accuracy and robustness in multipath environments.

\begin{figure}[t!]
\centering
\includegraphics[width =3.3in]{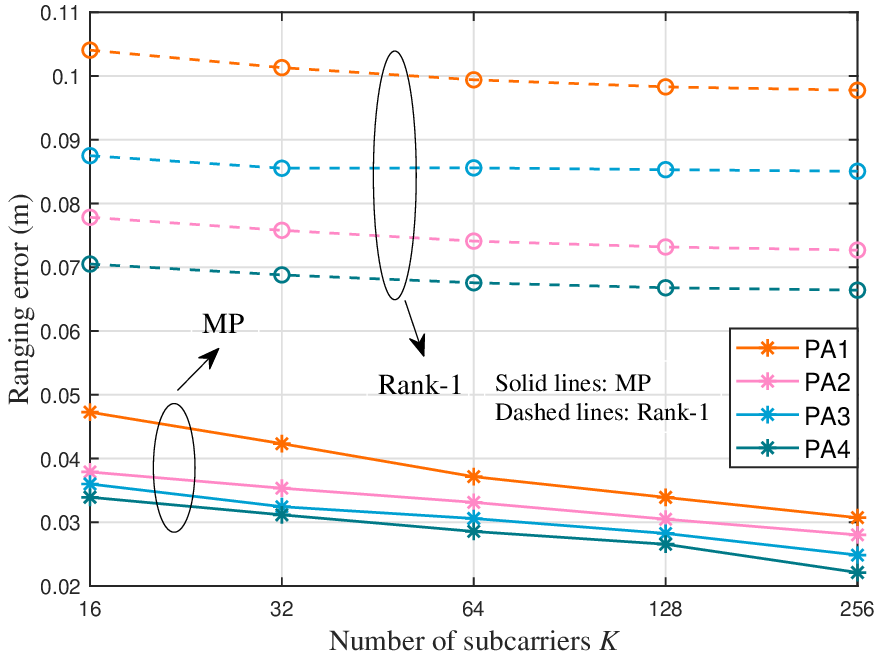}
\caption{The impact of the number of subcarriers on ranging errors, where the noise power is set to -90 dBm.}
\label{ranging carrier}
\end{figure}

\emph{2) The impact of the number of subcarriers on ranging errors:} Fig.~\ref{ranging carrier} illustrates the impact of the number of subcarriers $K$ on the ranging errors of the MP and Rank-1 algorithms. The ranging errors of both algorithms decrease as $K$ increases, which is consistent with ~\textbf{Remark~\ref{remark1:subcarrier}}. Furthermore, the MP algorithm exhibits a more pronounced performance improvement, whereas the error curves of the Rank-1 algorithm decrease slightly as the number of subcarriers increases. The reason is that the Rank-1 algorithm suffers from residual multipath interference caused by the one-dimensional subspace approximation, and the resulting bias cannot be eliminated by increasing the number of subcarriers. In contrast, the MP algorithm separates the LoS path from the NLoS components, enabling it to more effectively exploit the additional frequency-domain observations for improved ranging accuracy.

\begin{figure}[t!]
\centering
\includegraphics[width =3.3in]{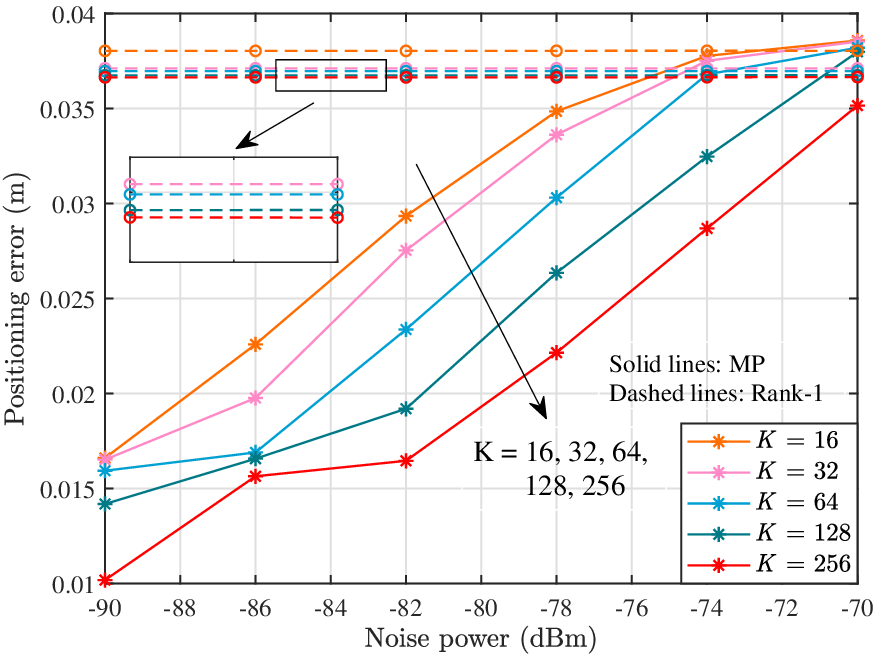}
\caption{Positioning errors of the MP-based and Rank-1-based WNLS algorithms under different numbers of subcarriers.}
\label{positioning error}
\end{figure}

\emph{3) Positioning errors based on MP and Rank-1 algorithms:} Fig.~\ref{positioning error} illustrates the positioning errors of the two-stage WNLS algorithm under varying noise power and subcarrier numbers $K$. We can see that the positioning error of the MP-based approach increases monotonically with the noise power, while the Rank-1-based approach exhibits the error floor across the evaluated noise range. Since the Rank-1-based approach is dominated by the deterministic NLoS bias in this noise range, the bias propagates through the WNLS estimator, resulting in a fixed positioning error floor that is insensitive to noise fluctuations. Furthermore, we can observe that increasing the number of subcarriers $K$ reduces the positioning error for the MP-based approach, but has a negligible impact on the Rank-1-based approach. This occurs because a larger number of subcarriers $K$ provides denser frequency-domain sampling, which effectively suppresses the ranging variance. For the MP algorithm, the positioning error is mainly governed by the noise power. Hence, increasing the number of subcarriers provides more frequency-domain observations and thus improves the positioning accuracy. In contrast, the Rank-1 algorithm is primarily limited by the NLoS-induced bias. Therefore, although a larger $K$ can reduce the random estimation variance, it cannot effectively suppress the multipath bias propagated from the ranging stage to the positioning stage. As the noise power increases to $-70$ dBm, the MP-based and Rank-1-based approaches achieve similar positioning accuracy, since random noise becomes the dominant error source. These results are consistent with the previous findings.

\begin{figure}[htbp]
\centering
\subfigure[MP-based positioning algorithm.]{\includegraphics[width =3.5in]{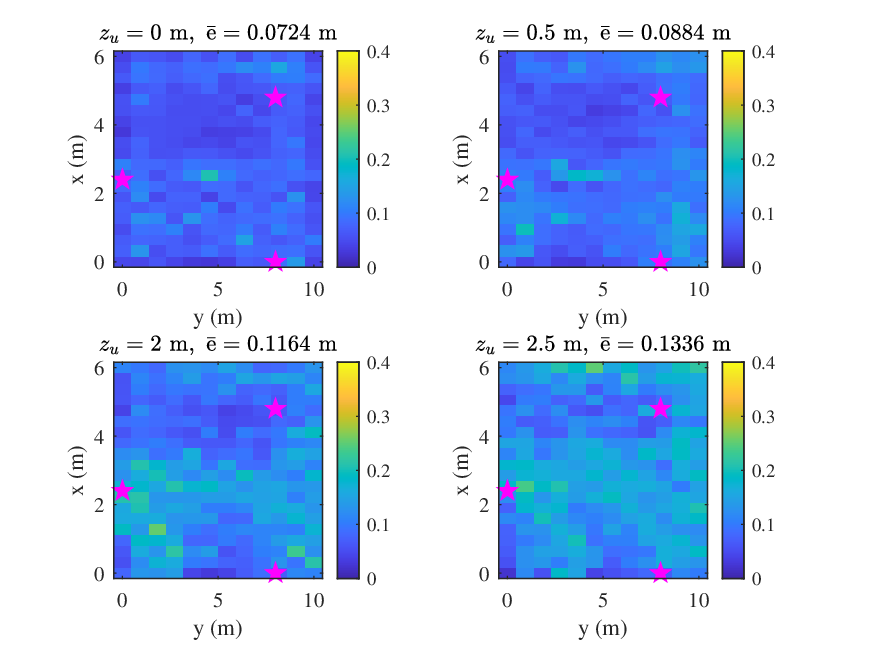} \label{MP maps}}
\subfigure[Rank-1-based positioning algorithm.]{\includegraphics[width =3.5in]{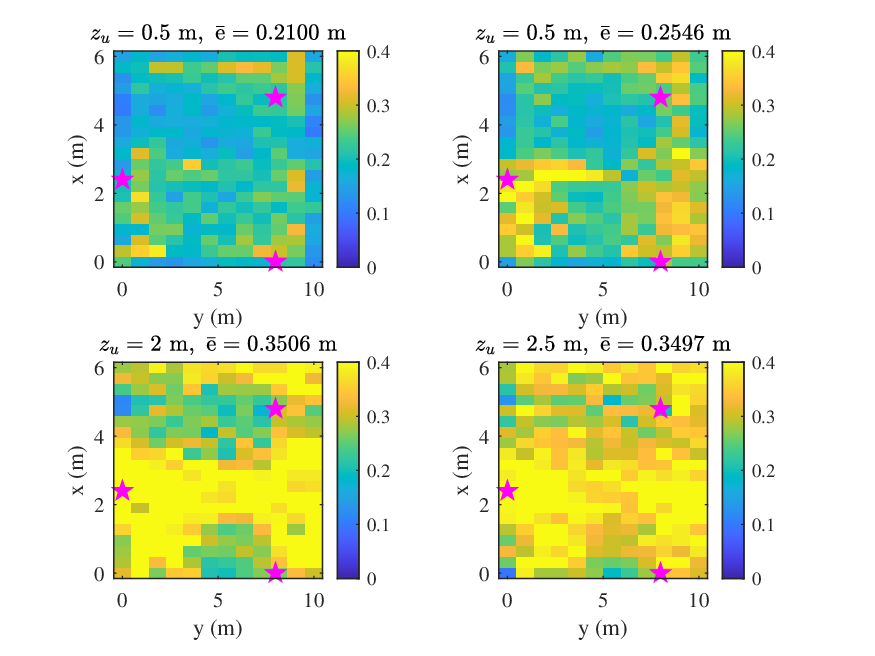} \label{R1 maps}}
\caption{Positioning error distributions, where the noise power and subcarrier number are set to -90 dBm and $K$ = 256, respectively.}
\label{positioning error distribution}
\end{figure}

\emph{4) Positioning error distribution:} Fig.~\ref{positioning error distribution} illustrates the spatial distribution of the positioning errors of the MP-based and Rank-1-based positioning algorithms under different user heights. Here, $\bar e$ denotes the average positioning error over the room, and the pink stars represent the PA locations. It can be observed that the MP-based algorithm achieves lower positioning errors over the entire room than the Rank-1-based algorithm. For both algorithms, the positioning errors are relatively uniformly distributed over the room. Moreover, the positioning error increases with the user height. This trend is more pronounced for the Rank-1-based algorithm, which demonstrates that the MP-based algorithm is more robust to user-position variations. For the MP-based algorithm, the average positioning error increases from $0.0724$ m to $0.1336$ m as the user height increases, while the error of the Rank-1-based algorithm increases from $0.2100$ m to $0.3497$ m. This is because the PA-user geometry varies with the user height. Since three of the four PAs are deployed on the ceiling, the user becomes closer to these ceiling-mounted PAs as the height increases, which leads to a larger GDoP and a smaller minimum eigenvalue of the range-based FIM. As a result, the same ranging error is amplified into a larger positioning error.

\begin{table*}[htbp]
\centering
\caption{Performance Comparison and Scenario Recommendations of the Proposed Algorithms}
\label{tab:algorithm_comparison}
\renewcommand{\arraystretch}{1.15}
\begin{tabular}{@{}>{\centering\arraybackslash}m{4cm}
                >{\centering\arraybackslash}m{4cm}
                >{\centering\arraybackslash}m{6cm}@{}}
\toprule
\textbf{Feature / Metric} & \textbf{MP-Based Algorithm} & \textbf{Rank-1-based Algorithm} \\ 
\midrule
\textbf{Extracted Information} & Full multipath information & LoS-related component only \\ 
\textbf{Estimator Property} & Nearly unbiased & Biased \\ 
\textbf{Accuracy} & High & Limited, but comparable to the MP algorithm when noise dominates \\
\textbf{Robustness} & Strong  & Limited \\  
\textbf{Computational Complexity} & ${\cal O}\left( {{\kappa ^3}} \right)$ & ${\cal O}\left( {{\kappa ^2}} \right)$ \\ 
\textbf{Pseudo-Inverse} & Required & Not required \\
\midrule
\textbf{Recommended Scenarios} & High-precision scenarios & Low-complexity or high-noise scenarios \\ 
\bottomrule
\end{tabular}
\end{table*}

\begin{figure}[t!]
\centering
\includegraphics[width =3.3in]{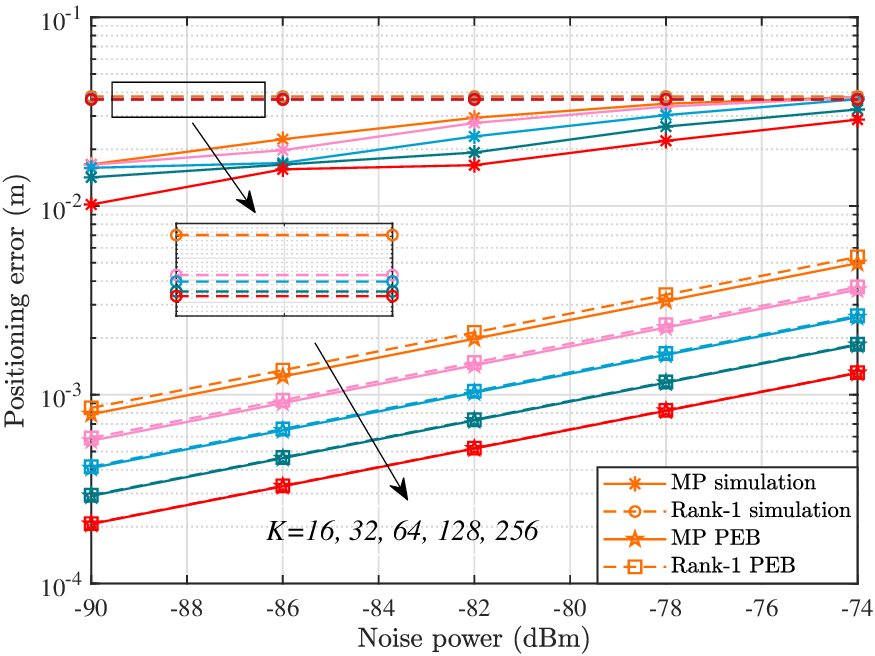}
\caption{The PEB of MP-based and Rank-1-based positioning algorithms.}
\label{PEB}
\end{figure}

\emph{5) The PEB of two-stage WNLS positioning algorithms:} Fig.~\ref{PEB} compares the positioning errors of the MP-based and Rank-1-based positioning algorithms with the derived PEB. It can be observed that the PEB increases with the noise power and decreases as the number of subcarriers increases, which is consistent with the theoretical analysis. This is because a lower noise power improves the effective SNR, while a larger number of subcarriers provides more frequency-domain observations for delay estimation. For the MP-based positioning algorithm, the simulated positioning errors follow the same trend as the PEB, which indicates that the dominant error source of the MP-based algorithm is well characterized by the PEB derivation. Since the MP algorithm separates the LoS path from the NLoS components, the residual deterministic multipath bias is relatively small. Therefore, its positioning performance is more consistent with the theoretical error bound. In contrast, the actual positioning error of the Rank-1-based approach deviates from the PEB, exhibiting an error floor across the noise range. The main reason is that the Rank-1 algorithm suffers from a deterministic bias induced by multipath interference. Under low-noise conditions, the positioning error is mainly dominated by this structural bias, making the theoretical variance bound unattainable. These results demonstrate that the MP algorithm can approach the theoretical limit more closely in multipath environments, whereas the Rank-1 algorithm trades positioning accuracy for lower computational complexity.

\emph{6) Summary and Scenario Recommendations:} Based on the above theoretical and simulation results, the two proposed algorithms exhibit different accuracy-complexity tradeoffs, as summarized in Table~\ref{tab:algorithm_comparison}. The MP-based algorithm estimates the full multipath information and separates the LoS component from the NLoS components, thereby achieving higher ranging and positioning accuracy. Its positioning error also follows the same trend as the derived PEB, indicating that the residual multipath bias is relatively small. However,  this performance gain is achieved at the cost of higher computational complexity, since the MP-based algorithm requires subspace extraction, pseudo-inverse computation, and eigenvalue decomposition. Therefore, the MP-based algorithm is more suitable for high-precision positioning scenarios where sufficient computational resources are available.

In contrast, the Rank-1-based algorithm extracts the LoS component through a Rank-1 Hankel approximation, avoiding pseudo-inverse computation and eigenvalue decomposition, which substantially reduces the computational complexity. However, this approximation introduces deterministic ranging bias, resulting in an error floor when the noise power is low. Therefore, the Rank-1-based algorithm is more suitable for low-complexity or high-noise scenarios.

\section{Conclusion}

In this paper, we established a comprehensive multi-carrier positioning framework for PASS in practical multipath environments. To accurately extract the multipath delays, we proposed the MP-based ranging algorithm to separate the LoS component from NLoS components. To further reduce the computational complexity, we developed a Rank-1 ranging algorithm that directly isolates the dominant LoS delay through truncated SVD. Based on the estimated LoS distances, we designed a two-stage WNLS algorithm to determine the 3D user position. To obtain deeper insights, we derived the closed-form ranging variances and the PEB to explicitly reveal the error propagation mechanism. Future work will investigate the positioning performance of PASS in multi-user scenarios.

\numberwithin{equation}{section}
\section*{Appendix~A: Proof of Lemma~\ref{lemma1:Eigenvalues}} \label{Appendix:As}
\renewcommand{\theequation}{A.\arabic{equation}}
\setcounter{equation}{0}

We multiply ${\bf H}_1$ by the pseudo-inverse of ${\bf H}_0$, denoted as ${\bf H}_0^\dagger$, which can be written as:
\begin{equation}\label{pseudo inverse}
\begin{aligned}
{{\bf{H}}_1}{\bf{H}}_0^\dag  &= ({{\bf{Z}}_M}{\bf{ZBZ}}_G^T){({{\bf{Z}}_M}{\bf{BZ}}_G^T)^\dag }\\
 &= {{\bf{Z}}_M}{\bf{ZBZ}}_G^T{({\bf{Z}}_G^T)^\dag }{{\bf{B}}^{ - 1}}{\bf{Z}}_M^\dag .
\end{aligned}
\end{equation}

By using the property of full column rank matrices, we can obtain the equations ${\bf{Z}}_G^T{({\bf{Z}}_G^T)^\dag } = {{\bf{I}}_L}$ and ${\bf{B}}{{\bf{B}}^{ - 1}} = {{\bf{I}}_L}$. Thus, \eqref{pseudo inverse} can be simplified to:
\begin{equation}\label{simplified H0H1}
\begin{aligned}
{{\bf{H}}_1}{\bf{H}}_0^\dag  = {{\bf{Z}}_M}{\bf{ZZ}}_M^\dag .
\end{aligned}
\end{equation}

Let ${\bf{P}} = {{\bf{Z}}_M}{\bf{Z}}$, and ${\bf{Q}} = {\bf{Z}}_M^\dag$. Then, we have:
\begin{equation}\label{QP}
\begin{aligned}
{\bf{QP}} = {\bf{Z}}_M^{^\dag }{{\bf{Z}}_M}{\bf{Z}} = {\bf{Z}}.
\end{aligned}
\end{equation}

According to matrix theory~\cite{matrix_analysis}, for the $L \times M$ matrix $\bf Q$ and $M \times L$ matrix $\bf P$, the non-zero eigenvalues of the product $\bf {PQ}$ are identical to those of $\bf {QP}$. Therefore, the $L$ non-zero eigenvalues of $ {{\bf{H}}_1} {\bf{H}}_0^\dag $ are exactly the diagonal elements of diagonal matrix $\bf Z$, which are the desired signal poles $\{z_{nl}\}_{l=1}^L$, and the proof is complete.

\numberwithin{equation}{section}
\section*{Appendix~B: Proof of Lemma~\ref{lemma2:delta}} \label{Appendix:Bs}
\renewcommand{\theequation}{B.\arabic{equation}}
\setcounter{equation}{0}

Let $N_0$ and $D_0$ denote the noise-free terms in the numerator and denominator, which can be written as:
\begin{equation}\label{noise-free terms}
N_0 = {z_{n1}}{\rm{Tr}}({({\bf{H}}_{n,A}^{\rm{o}})^H}{\bf{H}}_{n,A}^{\rm{o}}),~D_0 = {\rm{Tr}}({({\bf{H}}_{n,A}^{\rm{o}})^H}{\bf{H}}_{n,A}^{\rm{o}}).
\end{equation}

Let $\Delta N$ and $\Delta D$ denote the corresponding first-order perturbation terms, which can be given by:
\begin{equation}\label{first-order perturbation terms}
\begin{aligned}
\Delta N &={{\rm{Tr}}({{({\bf{H}}_{n,A}^{\rm{o}})}^H}{{\bf{N}}_{n,B}}) + {z_{n1}}{\rm{Tr}}({\bf{N}}_{n,A}^H{\bf{H}}_{n,A}^{\rm{o}})},\\
\Delta D &= {{\rm{Tr}}({{({\bf{H}}_{n,A}^{\rm{o}})}^H}{{\bf{N}}_{n,A}}) + {\rm{Tr}}({\bf{N}}_{n,A}^H{\bf{H}}_{n,A}^{\rm{o}})}.
\end{aligned}
\end{equation}

Then, applying the first-order Taylor expansion $(1+x)^{-1} \approx 1-x$ into \eqref{pole_expand}, the pole ${{\hat z}_{n1}}$ can be rewritten as:
\begin{equation}
\begin{aligned}
&{{\hat z}_{n1}} = \frac{{{N_0} + \Delta N}}{{{D_0} + \Delta D}} = \frac{{{N_0} + \Delta N}}{{{D_0}\left( {1 + \frac{{\Delta D}}{{{D_0}}}} \right)}}\\
& \approx \frac{{{N_0} + \Delta N}}{{{D_0}}}\left( {1 - \frac{{\Delta D}}{{{D_0}}}} \right) \approx \frac{{{N_0}}}{{{D_0}}} + \frac{{\Delta N - \frac{{{N_0}}}{{{D_0}}}\Delta D}}{{{D_0}}}.
\end{aligned}
\end{equation}

Consequently, the pole estimation error can be derived as follows:
\begin{equation}
\begin{aligned}
&\Delta {z_{n1}} = {{\hat z}_{n1}} - {z_{n1}} = \frac{{{N_0}}}{{{D_0}}} + \frac{{\Delta N - \frac{{{N_0}}}{{{D_0}}}\Delta D}}{{{D_0}}} - \frac{{{N_0}}}{{{D_0}}}\\
&= \frac{{\Delta N - {z_{n1}}\Delta D}}{{{D_0}}} = \frac{\mathrm{Tr}\big((\mathbf{H}_{n,A}^{\text{o}})^H(\mathbf{N}_{n,B} - z_{n1}\mathbf{N}_{n,A})\big)}{\|\mathbf{H}_{n,A}^{\text{o}}\|_F^2}.
\end{aligned}
\end{equation}

The proof is complete.

\numberwithin{equation}{section}
\section*{Appendix~C: Proof of Theorem~\ref{Theorem2:variance of alg2}} \label{Appendix:Cs}
\renewcommand{\theequation}{C.\arabic{equation}}
\setcounter{equation}{0}

Let the $N_{\mathrm{err}} = \mathrm{Tr}\big((\mathbf{H}_{n,A}^{\text{o}})^H (\mathbf{N}_{n,B} - z_{n1}\mathbf{N}_{n,A})\big)$. By expressing the trace operation as the sum of element-wise products, we have:
\begin{equation}\label{Nerr}
\begin{aligned}
N_{\mathrm{err}} = \beta_{n1}^* \sum_{i=0}^{M} \sum_{k=0}^{G-2} (z_{n1}^*)^{i+k} (w_{i+k+1} - z_{n1} w_{i+k})\\
=\beta_{n1}^* \sum_{l=0}^{M+G-2} c_l (z_{n1}^*)^l (w_{l+1} - z_{n1} w_l),
\end{aligned}
\end{equation}
where $c_l$ denotes the number of elements on the $l$-th anti-diagonal of the $(M+1) \times (G-1)$ matrix $\mathbf{H}_{n,A}^{\text{o}}$.

Then, the variance of the pole error is derived as:
\begin{equation}\label{variance of the pole error}
\mathrm{Var}(\Delta z_{n1}) = \frac{\mathrm{Var}(N_{\mathrm{err}})}{(\|\mathbf{H}_{n,A}^{\text{o}}\|_F^2)^2} = \frac{2\min(M+1, G-1)}{(M+1)^2(G-1)^2} \frac{\sigma_w^2}{|\beta_{n1}|^2}.
\end{equation}

For complex Gaussian noise, the variance of the phase error is half of the relative complex error variance, i.e., ${\rm{Var}}\left( {\arg \left( {{{\hat z}_{nl}}} \right)} \right) = \frac{1}{2}{\rm{Var}}\left( {\Delta {z_{n1}}} \right)$. Then, similar to Theorem~\ref{Theorem1:variance of alg1}, the variance in \eqref{variance of the pole error} can be mapped to the LoS distance-estimation variance in \eqref{var_d_rank1}, and the proof is complete.

\bibliographystyle{IEEEtran}
\bibliography{IEEEabrv,Uplink_Positioning_for_PASS_in_Multipath_Environments}

\end{document}